\documentclass{aa}  

\usepackage{graphicx}

\usepackage{txfonts}
\usepackage{natbib}
\bibpunct{(}{)}{;}{a}{}{,} 

\begin{document} 

   \title{Can circumbinary discs produce the eccentricities of shell-burning stripped giant binaries?}

   \author{C. A. S. Moltzer\inst{1},
          O. R. Pols\inst{1},
          H. Van Winckel\inst{2}
          }
    
   \institute{Department of Astrophysics/IMAPP, Radboud University Nijmegen, PO Box 9010, 6500 GL Nijmegen, The Netherlands\\
              \email{casper.moltzer@ru.nl}
         \and
             Institute of Astronomy, KU Leuven, Celestijnenlaan 200D, 3001 Leuven, Belgium\\
             }

   \date{Received X; accepted Y}

  \abstract
   {Post-red giant branch and post-asymptotic giant branch binaries, collectively shell-burning stripped giant (SBSG) binaries, contain a primary star that has recently been stripped of its envelope, alongside a main-sequence companion. These systems are characterised by a stable circumbinary disc (CBD), which is thought to have formed from the envelope material stripped via mass transfer. Their eccentricities range between $0-0.63$, contradicting canonical binary evolution, which predicts such post-mass-transfer systems should have circularised.}
   {We investigate whether interaction between a binary and its CBD can explain the observed eccentricities of SBSG binaries.}
   {We utilised a new formalism for CBD-binary interaction based on hydrodynamic simulations. To compare with the observations, we generated model populations for which the post-mass-transfer eccentricity and the amount of material accreted from the CBD were free parameters.}
   {We found that CBD-binary interaction can reproduce the observed eccentricity distribution of SBSG binaries, provided that: (1) their post-mass-transfer eccentricities range up to at least 0.05, (2) their CBDs have initial masses of at least $0.1$~$M_\odot$, and (3) accretion onto the SBSG star is highly inefficient in order to prevent refilling of its Roche lobe.}
   {Our model requires higher post-mass-transfer eccentricities than are canonically predicted and more massive CBDs than are currently observed. We speculate that there is a population of post-mass-transfer progenitor systems with CBDs massive enough to facilitate significant eccentricity pumping. Mass loss via the $L_2$ point during mass transfer needs to be investigated, as this could form the CBD and shorten the orbital period, which is necessary for many SBSG binaries assuming they formed stably. We hypothesise that the observed eccentricities are related to the amount of mass lost via $L_2$. Since many other post-interaction systems exhibit similar orbital properties, we speculate that they may all have had CBDs with which they interacted shortly after mass transfer.}


    \authorrunning{C. A. S. Moltzer et al.}
    \titlerunning{Can circumbinary discs produce the eccentricities of shell-burning stripped giant binaries?}
   \maketitle

\section{Introduction}
Post-asymptotic giant branch (post-AGB) binaries are systems in which the primary star has recently lost its envelope while on the AGB. This causes the star to contract to a higher effective temperature ($T_\mathrm{eff}$) while maintaining a constant luminosity through hydrogen shell burning \cite[e.g.][]{MillerBertolami2016}. Post-red giant branch (post-RGB) binaries are the RGB analogues of post-AGB binaries and are distinguished by luminosities below the tip of the RGB \citep[${\sim}2300$~$L_\odot$; e.g.][]{Kamath2015}. To date, 85 systems have been identified in the Milky Way and it is now well established that they are characterised by the presence of a stable circumbinary disc (CBD) of gas and dust \citep[e.g.][]{Kluska2022}. Additionally, candidate systems have been found in both the Large and Small Magellanic Clouds \citep[e.g.][]{vanAarle2011,Kamath2014,Kamath2015}. 

The orbital periods of post-RGB and post-AGB binaries range from 50 to 3000 days \citep[e.g.][]{Kluska2022}. This indicates that the envelopes of the primary stars were stripped via Roche-lobe overflow (RLOF), as their progenitor stars would not fit inside the Roche lobes of these orbits. Therefore, from a stellar evolutionary perspective, the post-RGB or post-AGB star in these binaries can be classified as a shell-burning stripped giant (SBSG). Henceforth, we will refer to these systems collectively as SBSG binaries, unless discussing post-RGB or post-AGB binaries separately. The companions of SBSG binaries are likely to be main-sequence stars \citep[e.g.][]{Oomen2018,Bollen2022}, indicating that these systems underwent a single mass transfer episode. While the orbital periods of post-RGB binaries are broadly consistent with stable RLOF \citep{Moltzer2025}, the majority of post-AGB binaries have considerably shorter orbital periods than predicted \citep[e.g.][]{Nie2012,VanWinckel2025}. Similarly to other low- and intermediate-mass post-interaction binary systems with orbital periods longer than ${\sim}100$ days \citep[e.g.][]{Shahaf2024,Mathieu2025}, SBSG binaries exhibit eccentric orbits. To date, 39 systems have had their eccentricities spectroscopically determined, with values ranging between 0 and 0.63 \citep[e.g.][]{Oomen2018}. This is at odds with the paradigm that these systems should have circularised post-mass-transfer orbits due to strong tidal forces before or during RLOF \citep[e.g.][]{Hurley2002}. 

In the case of giant star donors, it is not expected that the orbits will become completely circular following stable mass transfer. \citet{Phinney1992} pointed out that the gravitational quadrupole moment is altered stochastically by the convective eddies in the envelopes of these donors, imparting residual eccentricities between $10^{-7}$ and $10^{-3}$. While these residual eccentricities are consistent with those observed in millisecond pulsars with He white dwarf companions \citep[e.g.][]{Cohen2024}, formed by stable RLOF from an RGB donor to a neutron star, they cannot explain the substantially larger eccentricities observed in SBSG binaries. Several mechanisms have been proposed in the literature to pump the eccentricities of post-interaction binaries: interaction with a circumbinary disc \citep[e.g.][]{Waelkens1996,Vos2015}, phase-dependent mass loss \citep[e.g.][]{Soker2000,BonavicMarinovic2008}, phase-dependent RLOF \citep[e.g.][]{Sepinsky2007,Hamers2019,Parkosidis2026a}, white dwarf recoils \citep[e.g.][]{Izzard2010,ElBadry2018}, and dynamical interaction with a tertiary companion \citep[e.g.][]{Toonen2020}. As SBSG binaries are characterised by having circumbinary discs, CBD-binary interaction appears to be a plausible explanation, especially since there is evidence for ongoing interaction. Firstly, accretion of gas onto SBSG stars from their CBD is found to be a natural explanation for the photospheric depletion of refractory elements, which have condensed into dust and are therefore expected to remain in the CBD \citep[e.g.][]{Oomen2019}. Furthermore, SBSG binaries are observed to launch jets, which are thought to originate from the companion accreting material from the CBD \citep[e.g.][]{DePrins2024,VanWinckel2025}. 

Previous studies that investigated whether CBD-binary interaction could explain the observed eccentricities of SBSG binaries were inconclusive. Although \citet{Dermine2013} showed that resonances in the CBD induced by the binary were effective in pumping eccentricities, \citet{Rafikov2016} found that this mechanism was far less efficient, requiring massive, long-lived CBDs which they concluded were unlikely to occur. Building upon both works, \citet{Oomen2020} were unable to produce eccentricities greater than ${\sim}0.25$ via these resonances. Furthermore, while the model presented by \citet{Izzard2023} is able to reproduce the observed eccentricity of $0.2$ for V390~Vel with some modest adjustments, it is unclear whether it can effectively reach the larger eccentricities of other SBSG binaries.

More recent studies have performed hydrodynamic simulations of CBD-binary interaction to investigate its impact on binary orbits via gravitational and accretion torques \citep[e.g.][]{D'Orazio2021,Zrake2021,Siwek2023b}. Using a CBD-binary interaction formalism based on the hydrodynamic simulations by \citet{Siwek2023a,Siwek2023b}, \citet{Valli2024} found significant eccentricity pumping of up to ${\sim}0.5$ when at least $5-10\%$ of the initial binary mass was accreted. They concluded that the observed CBD masses of SBSG binaries, which typically range between $10^{-3}$ and $10^{-2}$~$M_\odot$ \citep[e.g.][]{GallardoCava2026}, are therefore too small to explain their eccentricities. However, this does not rule out the possibility that these CBDs were more massive in the past, at the time when the eccentricity pumping occurred. 

We use the formalism presented by \citet{Valli2024} to investigate whether CBD-binary interaction can account for the observed eccentricity distribution of SBSG binaries. In Sect.~\ref{methods}, we describe our CBD-binary interaction model. In Sect.~\ref{eccentricity}, we demonstrate how our model can reproduce the observed eccentricity distribution. We estimate the accretion efficiency of our model in Sect.~\ref{accretion efficiency}. In Sect.~\ref{orbital period}, we show the effect of CBD-binary interaction on orbital periods, assuming the systems formed via stable mass transfer. We discuss our model in Sect.~\ref{discussion}, and summarise our conclusions in Sect.~\ref{conclusion}.

\section{Methods} \label{methods}
In order to investigate whether CBD-binary interaction can produce the observed eccentricity distribution of SBSG binaries, we simulated binary systems undergoing CBD-binary interaction using the formalism presented by \citet{Valli2024}, as described in Sect.~\ref{disc-binary interaction}. We generated model populations of binaries with initial properties sampled from the observed systems (see Sect.~\ref{population_synthesis}). We statistically tested the CBD-binary interaction hypothesis by comparing the eccentricity distributions of the model population and the observed systems (see Sect.~\ref{ConTEST}).

\subsection{CBD-binary interaction formalism} \label{disc-binary interaction}
We used the formalism presented by \citet{Valli2024} to compute the evolution of a binary system undergoing long-term interaction with a CBD. This formalism uses the following three differential equations to describe the changes in orbital separation ($a$), eccentricity ($e$), and mass ratio ($q$): 
\begin{equation}
   \frac{d \log{a}}{d \log{M}}=f_a(e,q),
    \label{equation:da}
\end{equation}
\begin{equation}
   \frac{d e}{d \log{M}}=f_e(e,q),
    \label{equation:de}
\end{equation}
\begin{equation}
   \frac{d q}{d \log{M}}=f_q(e,q),
    \label{equation:dq}
\end{equation}
where $q \equiv M_\mathrm{s}/M_\mathrm{p} \leqslant 1$, with $M_\mathrm{p}$ and $M_\mathrm{s}$ being the primary and secondary masses, respectively; $M=M_\mathrm{p}+M_\mathrm{s}$ is the total binary mass; $f_a(e,q)$, $f_e(e,q)$, and $f_q(e,q)$ are functions that depend on $e$ and $q$. These functions are interpolated from the hydrodynamic simulations of \citet{Siwek2023a,Siwek2023b} on which this formalism is based. As $a$ and $M$ are scale-invariant in this formalism as a result of the hydrodynamic simulations being set up in this way, they can be written as logarithmic derivatives in Eqs.~\ref{equation:da}-\ref{equation:dq}. Consequently, the only free parameters within this formalism are $e$ and $q$. However, expressing the derivatives in terms of $M$ instead of time in order to achieve scale invariance means that information on the time dependence is no longer available. 

The study by \citet{Siwek2023a} investigated the preferential accretion rate of CBD-binary interaction onto the secondary compared with the primary, since this accretion has been found to evolve the mass ratios of binaries towards unity \citep[e.g.][]{Artymowicz1983,Duffell2020}. They performed 2D hydrodynamic simulations of a central binary, consisting of two sink particles, accreting gas from a finite, Keplerian-rotating, locally isothermal viscous CBD in a coplanar prograde orbit. Each simulation was integrated over 10000 orbital periods ($P_\mathrm{orb}$), and the rates of change were defined as the average over this duration. The simulations make the following assumptions: the radii of the stars are much smaller than their separation; the CBD mass is much smaller than the binary mass (i.e. the self-gravity of the CBD can be ignored); the change in orbital parameters takes much longer than the viscous relaxation time of the inner CBD region (i.e. the CBD is viscously relaxed); there are no outflows from the CBD or other mass-loss mechanisms, apart from accretion. Using the relative accretion rate $\lambda(e,q)=\dot{M_\mathrm{s}}/\dot{M_\mathrm{p}}$ tabulated in \citet{Siwek2023a}, $f_q(e,q)$ is computed by
\begin{equation}
   f_q(e,q)=\frac{(1+q)(\lambda(e,q)-q)}{1+\lambda(e,q)}.
\end{equation}

The study by \citet{Siwek2023b} investigated the change in $e$ and $a$ as a result of gravitational and accretion torques during CBD-binary interaction by conducting a series of hydrodynamic simulations similar to those by \citet{Siwek2023a}. The resulting $f_a(e,q)$ and $f_e(e,q)$ are tabulated in \citet{Valli2024}, which includes an additional suite of simulations obtained using the same methodology as by \citet{Siwek2023b}. These simulations reveal that $e$ evolves towards an equilibrium value $e_\mathrm{eq} \sim 0.5$, which depends slightly on $q$. This equilibrium value can be seen as a consequence of the eccentricity-pumping mechanism of CBD-binary interaction, since $e$ decreases or increases when the secondary is either accelerated or decelerated, respectively, as it approaches the inner edge of the CBD at apastron due to the difference in angular velocities \citep[e.g.][]{Artymowicz1991,Roedig2011}. Intuitively, $e_\mathrm{eq}$ can be interpreted as the point at which the angular velocity of the secondary at apastron is equal to that of the fluid at the inner edge of the CBD, although this is an oversimplification since $e_\mathrm{eq}$ represents the full non-linear solution resulting from the complex morphology of the CBD in the hydrodynamic simulations by \citet{Siwek2023b}.

We solved the three differential equations to simulate CBD-binary interaction by employing the same numerical methods as \citet{Valli2024}. When we interpolate the values for $f_a(e,q)$, $f_e(e,q)$, and $f_q(e,q)$, if $e$ and $q$ are outside of the defined range, the value of the nearest known data point is returned.

The hydrodynamic simulations tracked accretion of material onto the sink particles of the binary components. However, they did not model accretion onto the stars themselves, as the sink particles have larger radii than the stars they represent \citep{Siwek2023a,Siwek2023b}. This means that, although mass is accreted onto the sink particle, the star can still eject matter. This is actually observed to occur in the form of jets from the secondary star in many SBSG binaries \citep[e.g.][]{DePrins2024,VanWinckel2025}. Therefore, neither the simulations nor the formalism we use can be assumed to determine the accretion efficiency onto the stars themselves. Furthermore, in the case of significant accretion, we expect SBSG stars to regain their envelopes and to fill their Roche lobes again. Therefore, unless otherwise specified, we modelled the accretion onto the stars as fully inefficient. We further discuss the motivation behind this assumption and its effects on CBD-binary interaction in Sect.~\ref{accretion efficiency}. To model fully inefficient accretion, we set $f_q(e,q)=0$, since a lack of accreted material on either star means there is no change in mass ratio. We therefore regard the integration variable $M$ in Eqs.~\ref{equation:da}-\ref{equation:dq} as representing the mass flowing from the CBD towards the binary without necessarily being accreted by the stars themselves.

\subsection{Generating model populations} \label{population_synthesis}
The following parameters must be defined to solve the differential equations of the CBD-binary interaction formalism: the amount of accreted mass $\Delta M$, the eccentricity at the beginning of CBD-binary interaction $e_\mathrm{b}$, and the mass ratio at the beginning of CBD-binary interaction $q_\mathrm{b}$. Due to the scale invariance, the orbital separation at the beginning of CBD-binary interaction $a_\mathrm{b}$ can be set to 1.

Therefore, in order to generate a population of SBSG binaries that have undergone CBD-binary interaction, we need to sample $q_\mathrm{b}$, $e_\mathrm{b}$, and $\Delta M$. We constructed different model populations, each comprising 1000 sampled systems. First, we obtained $q_\mathrm{b}$ by sampling the SBSG mass ($M_1$) and companion mass ($M_2$) distributions, where $q_\mathrm{b} \equiv M_1/M_2$. Note that $q_\mathrm{b}$ differs from $q$ in Eqs.~\ref{equation:da}-\ref{equation:dq}: while $q=q_\mathrm{b}$ when $M_1 \leq M_2$, $q=1/q_\mathrm{b}$ when $M_1>M_2$. We sampled the $M_1$ and $M_2$ distributions from current observations since, under the assumption of fully inefficient accretion, these distributions remain unchanged by CBD-binary interaction. 

The $M_1$ distribution was derived from the observed SBSG binary luminosity distribution and converted using appropriate mass-luminosity relations. Since the contribution of the companion to the total luminosity is negligible \citep[e.g.][]{Kluska2022}, we assumed that the luminosity of the binary is equal to that of the primary. For post-RGB stars, we used the models of \citet{Moltzer2025} to determine the mass-luminosity relation (see Appendix~\ref{pRGB ML} for further details). For post-AGB stars, we used the models of \citet{MillerBertolami2016} and performed a first-degree polynomial fit to obtain
\begin{equation}
   \frac{L}{10^4 L_\odot}  =  6.4\pm0.2 \frac{M_1}{M_\odot} -3.1\pm0.1.
    \label{equation:ML}
\end{equation}
This relation applies to post-AGB stars formed via the single-star formation channel, and we ignored the spread that is expected around this relation for post-AGB stars resulting from binary interaction \citep[e.g. see the luminosities of the post-AGB models in Fig.~4 in][]{Moltzer2025}.

We assumed that all objects with a luminosity below the RGB-tip luminosity (${\sim}2300$ $L_\odot$) are post-RGB stars. As discussed by \citet{Moltzer2025}, although post-AGB stars with luminosities below the RGB-tip can form via RLOF, they are expected to be considerably less abundant than post-RGB stars. This is primarily because such post-AGB stars have higher initial masses (${\geqslant}2.0$ $M_\odot$), which are less abundant according to the initial mass function \citep[IMF; e.g.][]{Kroupa2001}. 

\begin{figure}
    \resizebox{\hsize}{!}{\includegraphics{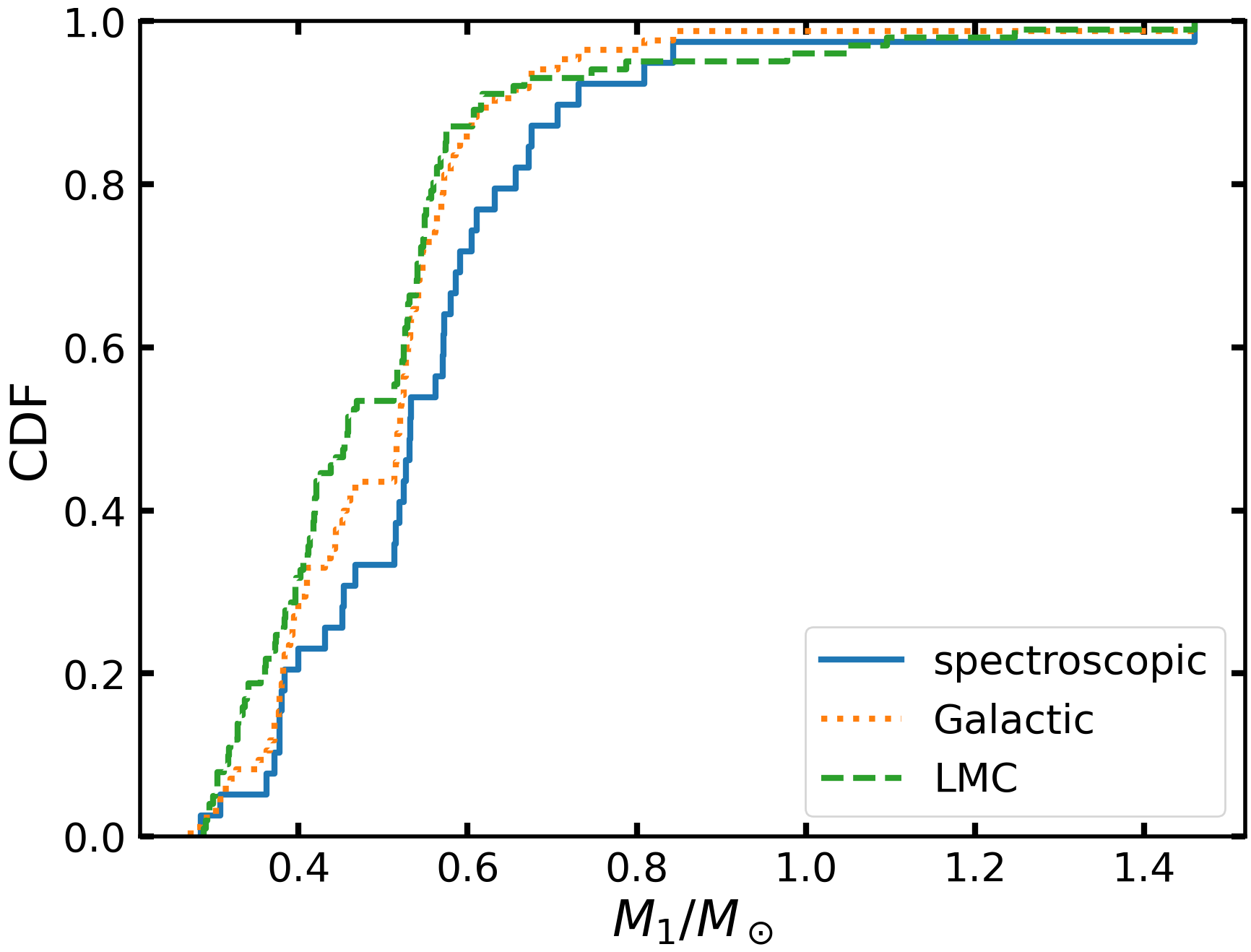}}
    \caption{Cumulative mass distribution of different samples of observed SBSG binary systems: the Galactic sample with spectroscopically determined orbits, the full Galactic sample, and the LMC candidate sample.}
    \label{figure:M_distribution_data}
\end{figure}

The cumulative mass distributions derived from the observed luminosity distributions taken from \citet{Moltzer2025} are shown in Fig.~\ref{figure:M_distribution_data} for the Galactic SBSG binaries with spectroscopically determined orbits, the full Galactic sample, and the Large Magellanic Cloud (LMC) candidate sample. The spectroscopic sample, and to a lesser extent the full Galactic sample, appears to contain a greater number of post-AGB stars ($M_1 > 0.47$~$M_\odot$) relative to post-RGB stars ($M_1 \leq 0.47$~$M_\odot$) than the LMC sample. However, the LMC sample is expected to be more complete since it is the result of a magnitude-limited systematic survey of the entire dwarf galaxy \citep{vanAarle2011,Kamath2015}. Therefore, we used the LMC luminosity distribution to sample $M_1$ from.

For each value of $M_1$, we subsequently selected a $M_2$ value by sampling the companion mass distribution of SBSG binaries presented by \citet{Oomen2018}, which is approximated by a normal distribution with a mean of 1.09 $M_\odot$ and a standard deviation of 0.62 $M_\odot$, in order to find $q_\mathrm{b}$. We restricted this distribution to the range $0.08-8.0$ $M_\odot$, i.e. the initial mass range of low- and intermediate-mass stars. We employed four different models of increasing complexity to sample the values for $\Delta M$ and $e_\mathrm{b}$, as described below.

\subsubsection{Model $\#1$}
Model $\#1$ is an exploratory model in which we used fixed values of $\Delta M$ and $e_\mathrm{b}$ ranging between $0.05-0.5$ $M_\odot$ and $0.01-0.1$, respectively. The $\Delta M$ range is much larger than the current CBD masses determined for SBSG binaries \citep[e.g.][]{GallardoCava2026}, while the $e_\mathrm{b}$ range is higher than the predicted residual eccentricities at the end of stable mass transfer \citep{Phinney1992}. This was done because we expect eccentricity pumping to be insufficient for such canonical values. To verify this, we also investigated model populations with $\Delta M$ fixed at the current CBD mass limit for SBSG binaries of $10^{-2}$ $M_\odot$, and with $e_\mathrm{b}$ fixed at the maximum eccentricity of $10^{-3}$ predicted by \citet{Phinney1992}.

\subsubsection{Model $\#2$}
It is thought that CBDs in SBSG binaries form from the envelopes of the donors, which are stripped during mass transfer. For model $\#2$, therefore, rather than using a fixed value for $\Delta M$, we assumed that $\Delta M$ is a fixed fraction $\delta$ of the envelope mass of the donor star before mass transfer. We ranged $\delta$ between $0.01-0.5$. The envelope mass was defined as the difference between the post-mass-transfer donor mass ($M_1$) and the pre-mass-transfer mass of the donor ($M_\mathrm{1,i}$). 

Although the $M_\mathrm{1,i}$ values are unknown a priori, they must be sufficiently high for the donor star to have been able to reach a core mass equal to $M_1$ at the onset of mass transfer. For single stars, there is a relationship between the initial mass of a low-mass star and its final mass after it sheds its envelope during the AGB phase to become a white dwarf; this is known as the initial-final mass relation \citep[IFMR; e.g.][]{Cummings2018}. This means that the IFMR provides the maximum core mass that a star of a given initial mass can reach during its evolution. Therefore, assuming negligible wind mass loss from the donor star prior to mass transfer, the IFMR provides a theoretical lower limit to the $M_\mathrm{1,i}$ value for a given $M_1$, since the envelope of the donor star may be stripped earlier in its evolution via mass transfer. 

Consequently, once we sampled $M_1$ for a model system, we determined $M_\mathrm{1,i}$ by sampling the IMF given by \citet{Kroupa2001} between a minimum and maximum value. The minimum mass was determined using the IFMR presented by \citet{Cunningham2024}\footnote{We estimated the IFMR by interpolating between the values provided in Table 2 of \citet{Cunningham2024}. We supplemented these values with the initial mass values of 0.9 and 8.0 $M_\odot$, and corresponding final mass values of 0.5464 and 1.40 $M_\odot$ computed using Eqs.~1 and 4 from \citet{Cunningham2024}, respectively.}, which was derived from a volume-limited sample of white dwarfs. The maximum mass was assumed to be either $2.0$ or $8.0$ $M_\odot$, depending on whether $M_1$ corresponded to a post-RGB or post-AGB star, respectively.

We limited $\Delta M$ to a maximum value of 1 $M_\odot$. Although CBD formation is currently not well understood, CBD masses greater than ${\sim}1$ $M_\odot$ are not supported by the amount of circumstellar material observed in systems studied by \citet{Khouri2021,Khouri2025}, which we speculate are similar to post-mass-transfer progenitor systems of SBSG binaries (see Sect.~\ref{disc masses}).

\subsubsection{Model $\#3$}
For model $\#3$, we imposed more stringent constraints on the companion mass. As we assigned the donor an initial mass $M_\mathrm{1,i}$ in model $\#2$, we can calculate the pre-mass-transfer mass ratio $q_\mathrm{i} \equiv M_\mathrm{1,i}/M_2$, where we equate the companion mass ($M_2$) before and after mass transfer by assuming negligible accretion during this process. The values of $q_\mathrm{i}$ must be greater than 1, as the more massive component will evolve into a giant star first, thus initiating mass transfer. Furthermore, we expect stable mass transfer to have formed these binaries, as their orbital periods are too long to be the result of common envelope evolution. To ensure stable mass transfer, $q_\mathrm{i}$ cannot exceed a critical value $q_\mathrm{crit}$ \citep[see e.g.][]{Temmink2023}, which we set at 2 for simplicity. Therefore, we assume that $M_2$ is distributed within the interval $[0.5 M_\mathrm{1,i}, M_\mathrm{1,i}]$.

Additionally, due to the negligible contribution of the companion to the total luminosity of the binary \citep[e.g.][]{Kluska2022}, we assumed a maximum photospheric contribution of $5\%$ by the companion. Using a zero-age main-sequence mass-luminosity relation\footnote{We estimated the zero-age main-sequence mass-luminosity relation using initially non-rotating models of solar metallicity taken from the MESA Isochrones $\&$ Stellar Tracks database \citep{Choi2016}.}, we derived another upper limit for $M_2$. We utilised the smallest of these upper limits when sampling the $M_2$ distribution.

\subsubsection{Model $\#4$}
It is unlikely that the eccentricity at the end of mass transfer will be the same for all binaries. In fact, the model of \citet{Phinney1992} predicts a stochastic distribution of post-mass-transfer eccentricities, which is consistent with the range of eccentricities observed in binary millisecond pulsars with He white dwarf companions \citep[e.g.][]{Cohen2024}, although these eccentricities are much smaller than those required by our models.

Therefore, for model $\#4$, rather than using a fixed value for $e_\mathrm{b}$, we sampled from a distribution. As there is no evidence to support a more specific choice, we used a uniform distribution between 0 and a maximum initial eccentricity $e_\mathrm{b,max}$, with $e_\mathrm{b,max}$ ranging between $0.001-0.1$.

\subsection{Comparing models to observations} \label{ConTEST}
We compared the eccentricity distributions of our synthesised populations with the observed SBSG binary distribution. For this, we used the sample of 39 objects with spectroscopically determined orbits, taking the eccentricities from Table C.1 in \citet{Moltzer2025}. For this comparison, we used a statistical consistency test presented by \citet{Stoppa2023} called \texttt{contest$\_$dens} (see Appendix~\ref{ConTEST details} for further details). We label models with $p$-values computed using this method that are greater than 0.05 as statistically significant. We found that in the vast majority of cases where \texttt{contest$\_$dens} computed a statistically significant $p$-value for a model population, a standard Kolmogorov-Smirnov test also yielded a $p$-value greater than 0.05.

\section{Results: eccentricity pumping} \label{eccentricity}
\subsection{Model $\#1$}
\begin{figure}
    \resizebox{\hsize}{!}{\includegraphics{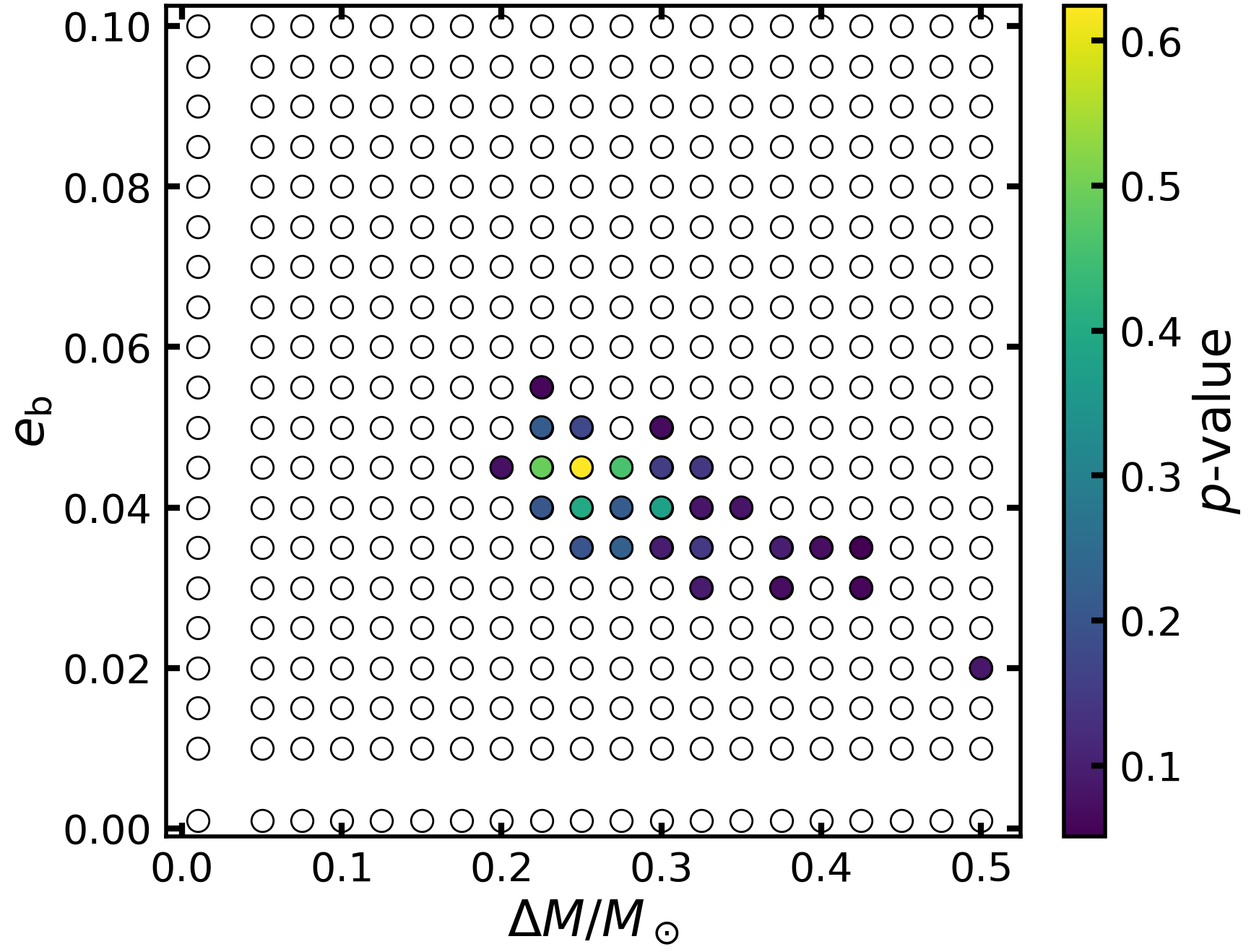}}
    \caption{Statistically significant $p$-values from the comparison of the eccentricity distributions of observed SBSG binaries and model $\#1$, for a range of $\Delta M$ and $e_\mathrm{b}$ values.}
    \label{figure:model1_results}
\end{figure}

Fig.~\ref{figure:model1_results} shows the $p$-values of model $\#1$ populations with different fixed values of $\Delta M$ and $e_\mathrm{b}$ above the significance value. The highest $p$-values are centred around $\Delta M=0.275$ $M_\odot$ and $e_\mathrm{b}=0.045$, but a valley of statistically significant distributions is found extending to $\Delta M=0.5$ $M_\odot$ and $e_\mathrm{b}=0.02$. 

\begin{figure*}
\centering
   \includegraphics[width=18cm]{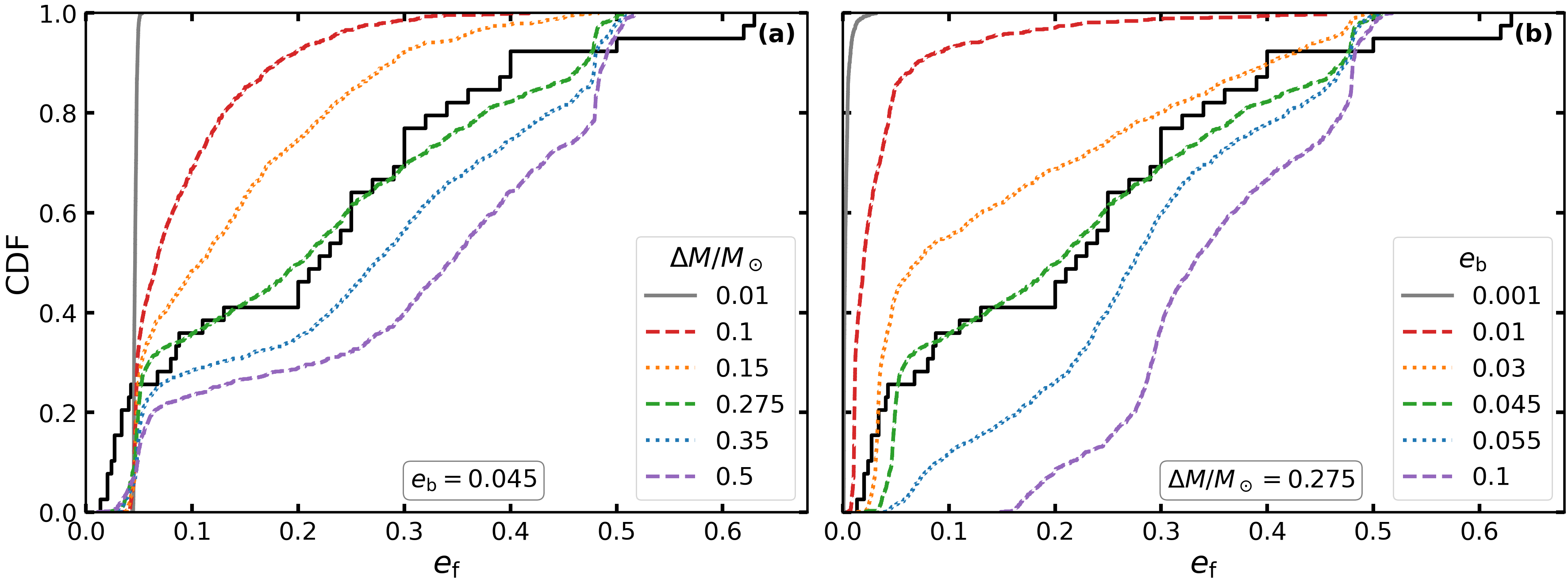}
     \caption{Cumulative eccentricity distributions of model $\#1$ populations with different values of $e_\mathrm{b}$ or $\Delta M$. Panel a shows model populations which all have $e_\mathrm{b}=0.045$ and panel b shows those which all have $\Delta M=0.275$~$M_\odot$. The solid black line in both panels denotes the observed eccentricity distribution of SBSG binaries. These eccentricities are defined as the median values of their corresponding truncated normal distributions (see Appendix~\ref{truncated median}).}
     \label{figure:model1_spread}
\end{figure*}

The reason for this relationship between $\Delta M$ and $e_\mathrm{b}$ can be seen in Fig.~\ref{figure:model1_spread}. A low value for either parameter results in a population with only small eccentricities. This is exemplified by model populations with $e_\mathrm{b}$ equal to the maximum eccentricity of ${\sim}10^{-3}$ predicted by \citet{Phinney1992} and those with $\Delta M$ equal to the observed upper limit of SBSG binary CBD masses (${\sim}10^{-2}$ $M_\odot$). As seen in Fig.~\ref{figure:model1_spread}, these model populations exhibit negligible eccentricity pumping. This demonstrates that larger values of either parameter are required for the CBD-binary interaction model to explain the observed SBSG binaries. Larger values of either $\Delta M$ or $e_\mathrm{b}$ result in an increase in high-eccentricity systems within a model population, with a secondary subpopulation forming around $e_\mathrm{eq}\sim0.5$. The valley of statistically significant model populations in the $\Delta M$ and $e_\mathrm{b}$ parameter space seen in Fig.~\ref{figure:model1_results} has the required proportion of low- and high-eccentricity systems to reproduce the observations. However, the $\Delta M$ and $e_\mathrm{b}$ values which give statistically significant model populations are significantly higher than the values expected from current observations and theory (see Sect.~\ref{discussion}). 

When comparing the eccentricity distribution of the model $\#1$ population with the highest $p$-value ($\Delta M=0.275$~$M_\odot$ and $e_\mathrm{b}=0.045$) with the observed distribution, shown in both panels of Fig.~\ref{figure:model1_spread}, three main differences emerge. Firstly, systems with the lowest eccentricities ($e<0.1$) exhibit less spread in the model, with the majority clumped between $0.0035-0.0055$. This is a consequence of the model having a fixed $e_\mathrm{b}$ for all systems. Secondly, the bimodal nature of the observed eccentricity distribution suggested by the gap between 0.13 and 0.20 \citep{Oomen2018} is not clearly evident in the model distribution. Thirdly, the two systems with the highest eccentricities ($e>0.5$) are not explained by the model. This is an inherent feature of all our models, as these eccentricities exceed the equilibrium eccentricity of ${\sim}0.5$ found in the utilised hydrodynamic simulations.

\subsection{Model $\#2$} 
\begin{figure}
    \resizebox{\hsize}{!}{\includegraphics{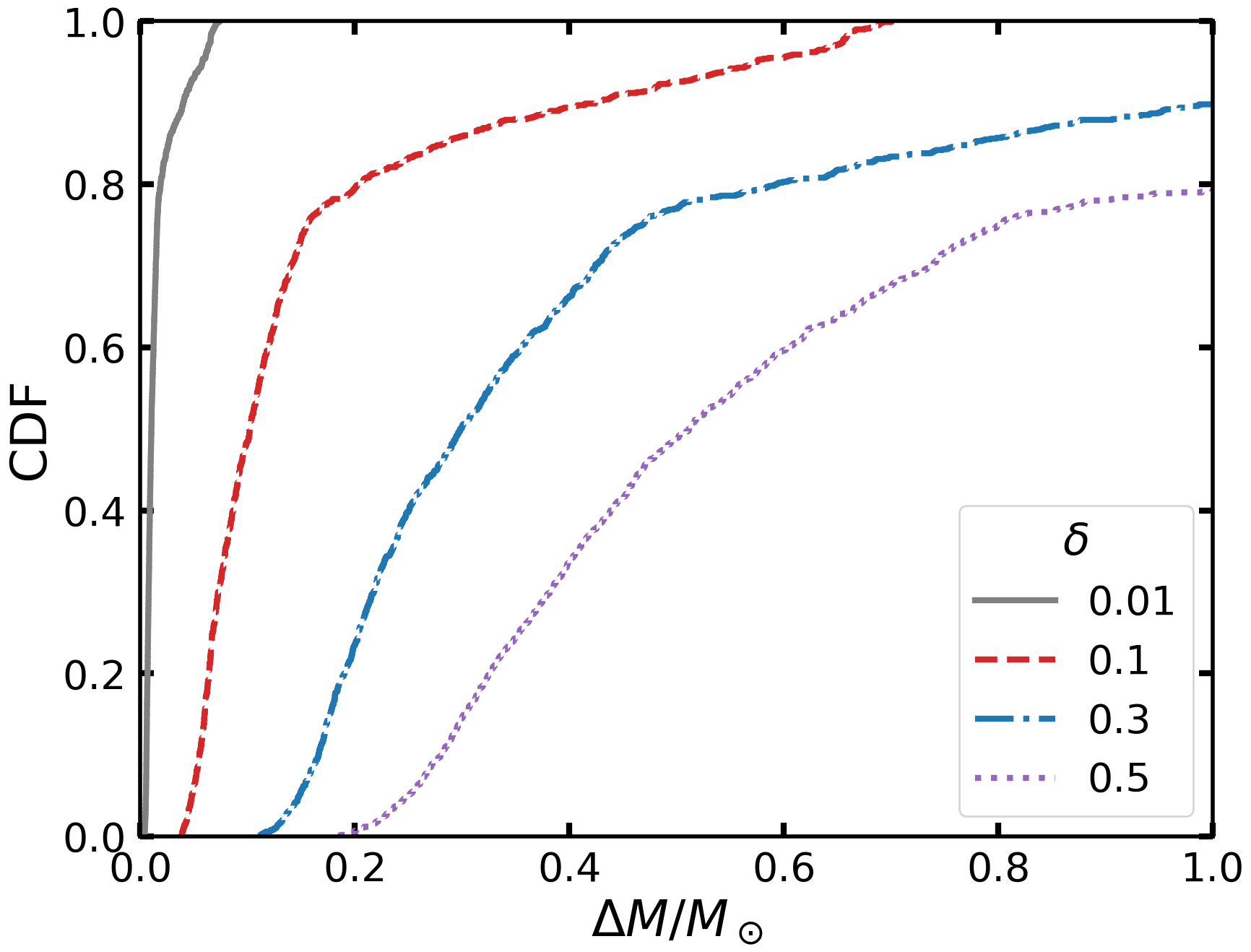}}
    \caption{Cumulative accreted mass distributions for different $\delta$ values. Note that $\Delta M$ is limited to 1.0 $M_\odot$.}
    \label{figure:model2_deltaM_distribution}
\end{figure}

In model $\#2$, we assume $\Delta M$ of each modelled system is a fixed fraction $\delta$ of the envelope mass stripped from the donor during mass transfer. This results in a distribution of $\Delta M$ for a model population that depends only on $\delta$. The key features of these $\Delta M$ distributions, shown in Fig.~\ref{figure:model2_deltaM_distribution}, are: a median $\Delta M/M_\odot$ value of ${\sim}\delta$, a minimum $\Delta M/M_\odot$ value of ${\sim}0.4\delta$, and a 75th percentile $\Delta M/M_\odot$ value of ${\sim}1.6\delta$. The upper 25th percentile extends to large $\Delta M$ values, reaching the imposed $\Delta M$ limit of 1.0 $M_\odot$ for $\delta \gtrsim 0.15$.

We find that model $\#2$ is able to reproduce the observed eccentricity distribution with similar $e_\mathrm{b}$ values to those of model $\#1$ and with $\delta$ values of ${\sim}0.2$. This $\delta$ value corresponds to a median $\Delta M$ value of 0.20 $M_\odot$, which is similar to the $\Delta M$ values for model $\#1$ that produce statistically significant model populations. The statistically significant model $\#2$ populations have fewer low-eccentricity systems and more systems around $e_\mathrm{eq}$ than statistically significant model $\#1$ populations. This is because the $\Delta M$ distributions in model $\#2$ have a substantial spread to both higher and lower values, which allows for systems within a population to experience either relatively much weaker or stronger eccentricity pumping.

\subsection{Model $\#3$} \label{model3_results}
\begin{figure}
    \resizebox{\hsize}{!}{\includegraphics{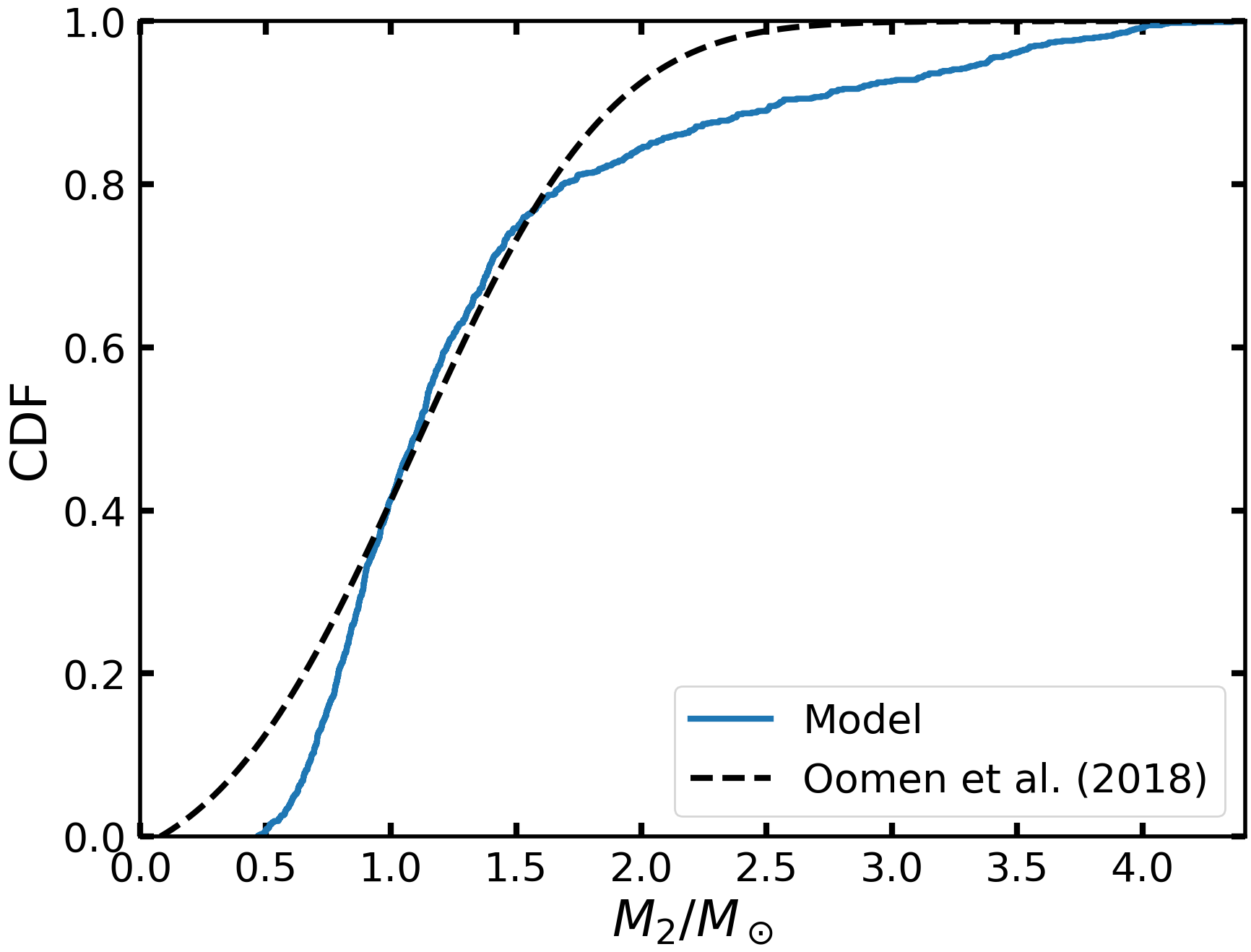}}
    \caption{Cumulative companion mass distributions of the model populations with the imposed constraint of $1 \leq q_\mathrm{i} \leq 2$, and the one presented by \citet{Oomen2018} from which these model populations were sampled.}
    \label{figure:model3_Mcomp_comparison}
\end{figure}

In model $\#3$, we constrained the pre-mass-transfer mass ratio $q_\mathrm{i}$ of each system to $1 \leq q_\mathrm{i} \leq 2$ to ensure that they were likely formed via stable mass transfer. This constraint results in a greater proportion of high-mass companions in the $M_2$ distribution, as shown in Fig.~\ref{figure:model3_Mcomp_comparison}. This distribution is identical for each model $\#3$ population and therefore differs from the companion mass distribution of \citet{Oomen2018}, from which these $M_2$ were originally sampled. However, as shown in Appendix~\ref{mass function}, the $M_2$ distribution of model $\#3$ cannot be statistically distinguished from the companion mass distribution of \citet{Oomen2018} based on the observed spectroscopic mass functions.

For model $\#3$, we find a valley of statistically significant populations centred around the highest $p$-value for $\delta=0.5$ and $e_\mathrm{b}=0.025$, and extending to $\delta=0.4$ and $e_\mathrm{b}=0.035$. This valley would extend to higher $\delta$ values if not for the maximum of 0.5 we imposed. However, we consider such high $\delta$ values to be implausible.

A major difference with model $\#2$ is that the $\delta$ values required for statistically significant populations are higher for model $\#3$. This results in much higher $\Delta M$ values as seen in Fig.~\ref{figure:model2_deltaM_distribution}, with the median shifting from 0.2 $M_\odot$ for model $\#2$ to 0.5 $M_\odot$. This is a direct result of the stable mass transfer constraint on $q_\mathrm{i}$, which removes systems with relatively low $M_2$. The larger $M_2$ values mean that the majority of $q_\mathrm{b}$ values in model $\#3$ are much smaller than in model $\#2$. As eccentricity pumping decreases in strength for small initial eccentricities when $q_\mathrm{b} \lesssim 0.6$, higher $\Delta M$ values are needed to reach the observed high-eccentricity systems.

\subsection{Model $\#4$}
\begin{figure}
    \resizebox{\hsize}{!}{\includegraphics{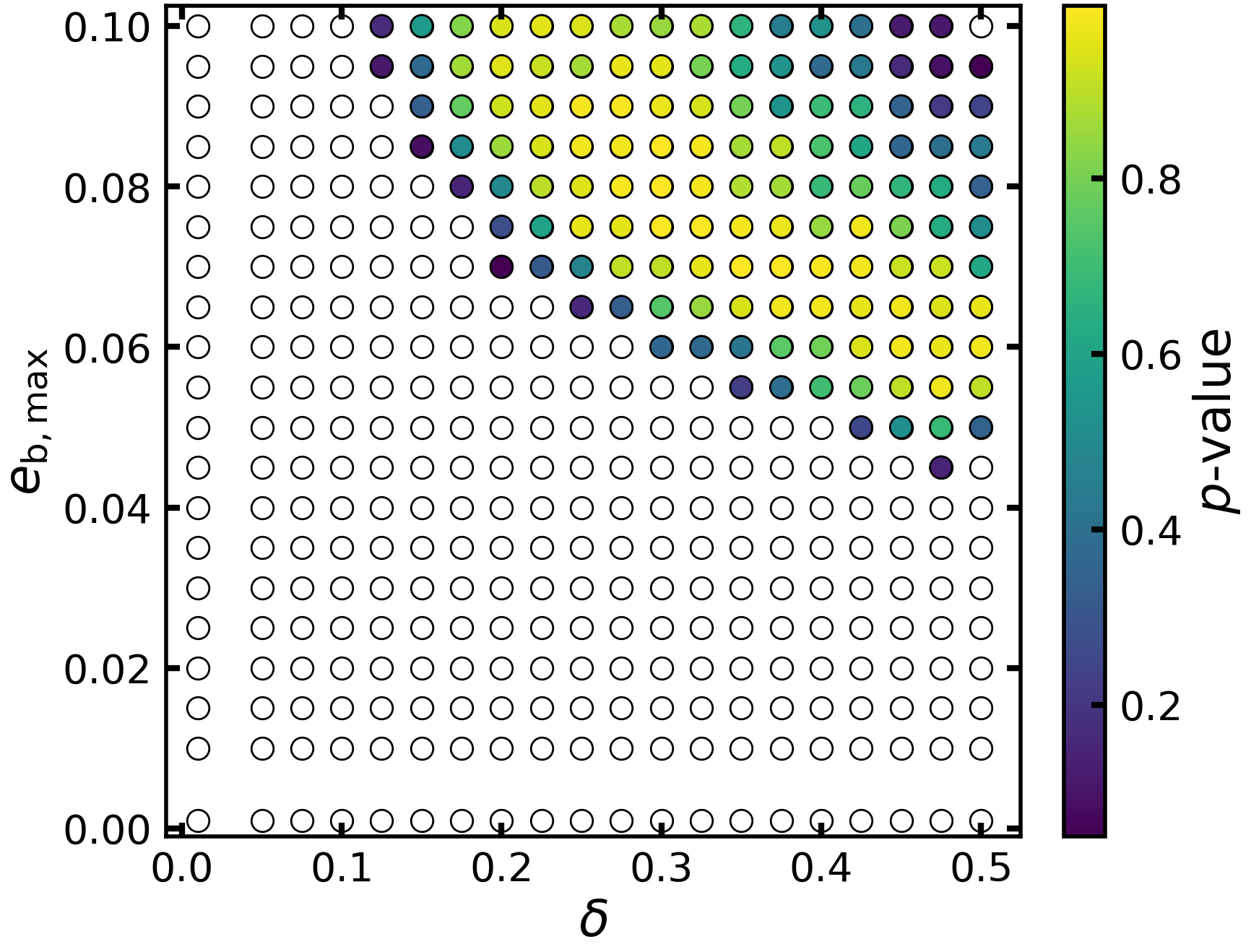}}
    \caption{Statistically significant $p$-values from the comparison of the eccentricity distributions of observed SBSG binaries and model $\#4$, for a range of $\delta$ and $e_\mathrm{b,max}$ values.}
    \label{figure:model4_results}
\end{figure}

In model $\#4$, rather than using a fixed $e_\mathrm{b}$ value for each system within the population, we sampled $e_\mathrm{b}$ from a uniform distribution between 0 and $e_\mathrm{b,max}$. Fig.~\ref{figure:model4_results} shows the $p$-values of model $\#4$ for different fixed values of $\delta$ and $e_\mathrm{b,max}$ above the significance value. The model population with the highest $p$-value has $\delta=0.35$ and $e_\mathrm{b,max}=0.07$, which is part of a broad valley of solutions ranging from $\delta=0.15$ and $e_\mathrm{b,max}=0.1$ to $\delta=0.5$ and $e_\mathrm{b,max}=0.05$. 

Compared to model $\#3$, model $\#4$ has a much larger parameter space that produces statistically significant populations. This means that less fine-tuning of these parameters is needed to reproduce the observed eccentricity distribution. While the $e_\mathrm{b,max}$ values required by model $\#4$ to produce the correct proportion of low- and high-eccentricity systems are higher than the $e_\mathrm{b}$ values needed by model $\#3$, the average $e_\mathrm{b}$ values are comparable to those of model $\#3$. Systems with $e_\mathrm{b}$ close to $e_\mathrm{b,max}$ are needed to compensate for systems with $e_\mathrm{b}$ values sampled between $0-0.01$, which experience negligible eccentricity pumping. On the other hand, this means that model $\#4$ allows for lower $\delta$ values, as these systems with a high $e_\mathrm{b}$ can be pumped much more easily and therefore require smaller $\Delta M$ values.

\subsection{Differences between post-RGB and post-AGB binaries} \label{pRGB vs pAGB}
\begin{figure}
    \resizebox{\hsize}{!}{\includegraphics{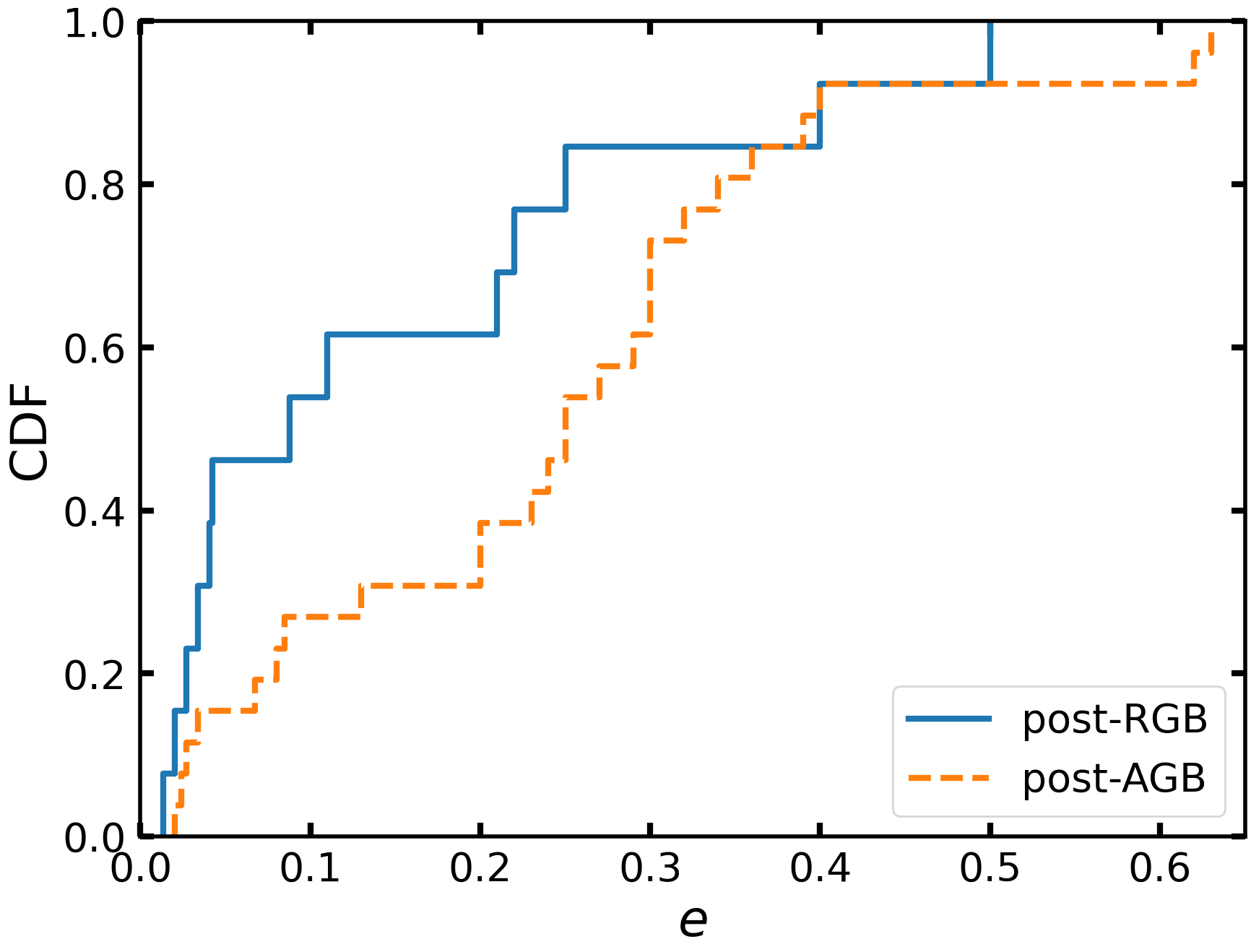}}
    \caption{Cumulative eccentricity distributions of the observed post-RGB and post-AGB binaries, individually. These eccentricities are defined as the median values of their corresponding truncated normal distributions (see Appendix~\ref{truncated median}).}
    \label{figure:e_distribution_separate}
\end{figure}

So far, we have treated the observed post-RGB and post-AGB binaries as a single population. However, they formed from donors that underwent binary interaction at different evolutionary stages. Consequently, they may have experienced CBD-binary interaction with different $e_\mathrm{b}$ and $\Delta M$ values. As mentioned in Sect.~\ref{population_synthesis}, we differentiate between the two object types by comparing the luminosities, determined through spectral energy distribution fitting by \citet{Moltzer2025}, to the RGB-tip luminosity of 2300 $L_\odot$. As seen in Fig.~\ref{figure:e_distribution_separate}, the observed eccentricity distributions of post-RGB and post-AGB binaries are noticeably distinct. Firstly, around $45\%$ of post-RGB binaries have very low eccentricities ($e<0.05$), whereas for post-AGB binaries this is only ${\sim}15\%$. Secondly, ${\sim}60\%$ of post-AGB binaries have eccentricities clustered around $0.3\pm0.1$, whereas such clustering is not apparent for post-RGB binaries. Lastly, only post-AGB binaries exhibit eccentricities higher than $e_\mathrm{eq}$. Although the distinction between post-RGB and post-AGB binaries is relatively uncertain due to the substantial errors on the luminosities, we find that removing the nine systems for which the error bars intersect the RGB-tip luminosity has an insignificant effect on the eccentricity distributions of both binary types.

In order to investigate these differences, we compared the post-RGB and post-AGB binary eccentricity distributions of the model $\#4$ populations with the observed distributions. We find that the majority of the statistically significant model populations shown in Fig.~\ref{figure:model4_results} for the combined sample are also significant for both the post-RGB and post-AGB binary populations when considered separately. However, for post-RGB binaries this valley of solutions extends to much lower values of  $e_\mathrm{b,max}$ (as low as 0.025 for $\delta=0.5$) and $\delta$ (as low as $10^{-2}$ for $e_\mathrm{b,max}=0.06$). This is a natural consequence of the greater number of systems with very low eccentricities. This suggests that post-RGB binaries could have lower $e_\mathrm{b}$ or $\Delta M$ compared to post-AGB binaries, a topic discussed further in Sect.~\ref{discrepancies}. 

\section{Results: accretion efficiency} \label{accretion efficiency}
Our CBD-binary interaction model populations that reproduce the observed eccentricity distribution of SBSG binaries require $0.1-1$~$M_\odot$ to be accreted from the inner edge of the CBD onto the binary. In the hydrodynamic simulations on which our model is based, the binary is represented as two sink particles on which this matter $\Delta M$ is accreted \citep{Siwek2023a,Siwek2023b}. However, if this amount of mass was accreted onto the stars themselves, we expect SBSG stars to regain their envelopes and enter another giant star phase. In this case, these stars would start filling their Roche lobes again and initiate another phase of mass transfer. We will show in the following that this can only be avoided if the accretion efficiency of the SBSG stars from the CBD is very small, which is why we assumed fully inefficient accretion in our model. We find that this assumption has a negligible effect on the eccentricity distribution following CBD-binary interaction, since the gravitational torques that govern eccentricity pumping remain unaffected \citep{Siwek2023b}; see Appendix~\ref{effect on orbital parameters} for further details.

To prevent a SBSG star from regaining its envelope through accretion, the accretion rate must be lower than the steady burning rate of the envelope $\dot{M}_\mathrm{burn}$. For these hydrogen shell-burning objects, the burning rate can be expressed as
\begin{equation}
  \dot{M}_\mathrm{burn} = \frac{L}{X_\mathrm{H} E_\mathrm{H}} \approx 1.4 \times 10^{-11} \frac{L}{L_\odot} \frac{M_\odot}{\mathrm{years}},
\end{equation}
where $X_\mathrm{H}\approx0.7$ is the abundance of hydrogen in the burning shell for stars with solar metallicity, and $E_\mathrm{H}\approx6.3\times10^{18}$~$\mathrm{erg}/\mathrm{g}$ is the energy released per unit mass of hydrogen \citep[e.g.][]{Kippenhahn2012}.

In order to assign accretion rates to the model systems, we need to consider the timescales $\tau$ during which CBD-binary interaction occurs. Detailed binary evolution models analysed by \citet{Moltzer2025} show that SBSG stars formed via stable mass transfer spend most of their evolutionary timescale at $T_\mathrm{eff}$ values lower than those observed in current systems. Therefore, we assume that the CBD-binary interaction timescales are approximately equal to the evolutionary timescales from the end of mass transfer to $T_\mathrm{eff}=4500$~K\footnote{Nearly all observed SBSG binaries have $T_\mathrm{eff}$ values greater than 4500~K. Since most of the evolutionary timescale is spent close to the $T_\mathrm{eff}$ value where mass transfer ends, the exact $T_\mathrm{eff}$ value that defines the end of the evolutionary timescale has a negligible effect on the computed $\tau$ value.}. Using the models from \citet{Moltzer2025}, we find that these evolutionary timescales follow approximately a power-law relationship with $L$: for post-RGB stars, between $\tau\approx1\times10^6$~years when $L\approx100$~$L_\odot$ and $\tau\approx5\times10^4$~years when $L\approx2300$~$L_\odot$; for post-AGB stars, between $\tau\approx6\times10^3$~years when $L\approx2300$~$L_\odot$ and $\tau\approx2\times10^2$~years when $L\approx1.1\times10^4$~$L_\odot$.

In the case of accretion onto a SBSG star, the evolutionary timescale is extended because the additional envelope mass results in slower evolution towards higher $T_\mathrm{eff}$ \citep[e.g][]{Oomen2018}. The study by \citet{Oomen2019}, which models the depletion of refractory elements observed in SBSG binaries through accretion from their CBDs, found that the evolutionary timescales of post-RGB stars and post-AGB stars are extended by a factor of up to two and five, respectively. We apply these factors to the aforementioned timescales. 

Finally, we can approximate the maximum possible accretion efficiency for each model system as $\epsilon_\mathrm{max} \equiv  \dot{M}_\mathrm{burn} \tau /\Delta M_1$, where $\Delta M_1$ is the portion of $\Delta M$ accreted onto the sink particle representing the SBSG star. For the model $\#4$ population with the highest $p$-value ($\delta=0.35$ and $e_\mathrm{b,max}=0.07$), we find that $\epsilon_\mathrm{max}$ is $5\times10^{-5}-10^{-2}$ for post-AGB stars and $7\times10^{-3}-4\times10^{-2}$ for post-RGB stars. These upper limits on the accretion rates are low enough for the accretion to be approximated as fully inefficient, as was done in our model. 

The assumption that no mass was accreted onto the main-sequence companion is supported by the observed outflows in the form of jets in SBSG binaries, which are believed to launch from the companions as a result of accretion from the CBD \citep[e.g.][]{DePrins2024,VanWinckel2025}. Although the efficiency with which the companion accretes matter is unknown, we expect its effect on CBD-binary interaction in terms of eccentricity pumping to be negligible, similarly to accretion onto the binary as a whole (see Appendix~\ref{effect on orbital parameters}).

\section{Results: orbital period change} \label{orbital period}
\subsection{Relative change in orbital period} 
Our CBD-binary interaction models are set up to be scale-invariant with respect to orbital separation. However, they do predict the relative change in the orbital separation, and consequently the change between the initial and final orbital period $P_\mathrm{orb,b}$ and $P_\mathrm{orb,f}$, as a result of the CBD-binary interaction.

\begin{figure}
    \resizebox{\hsize}{!}{\includegraphics{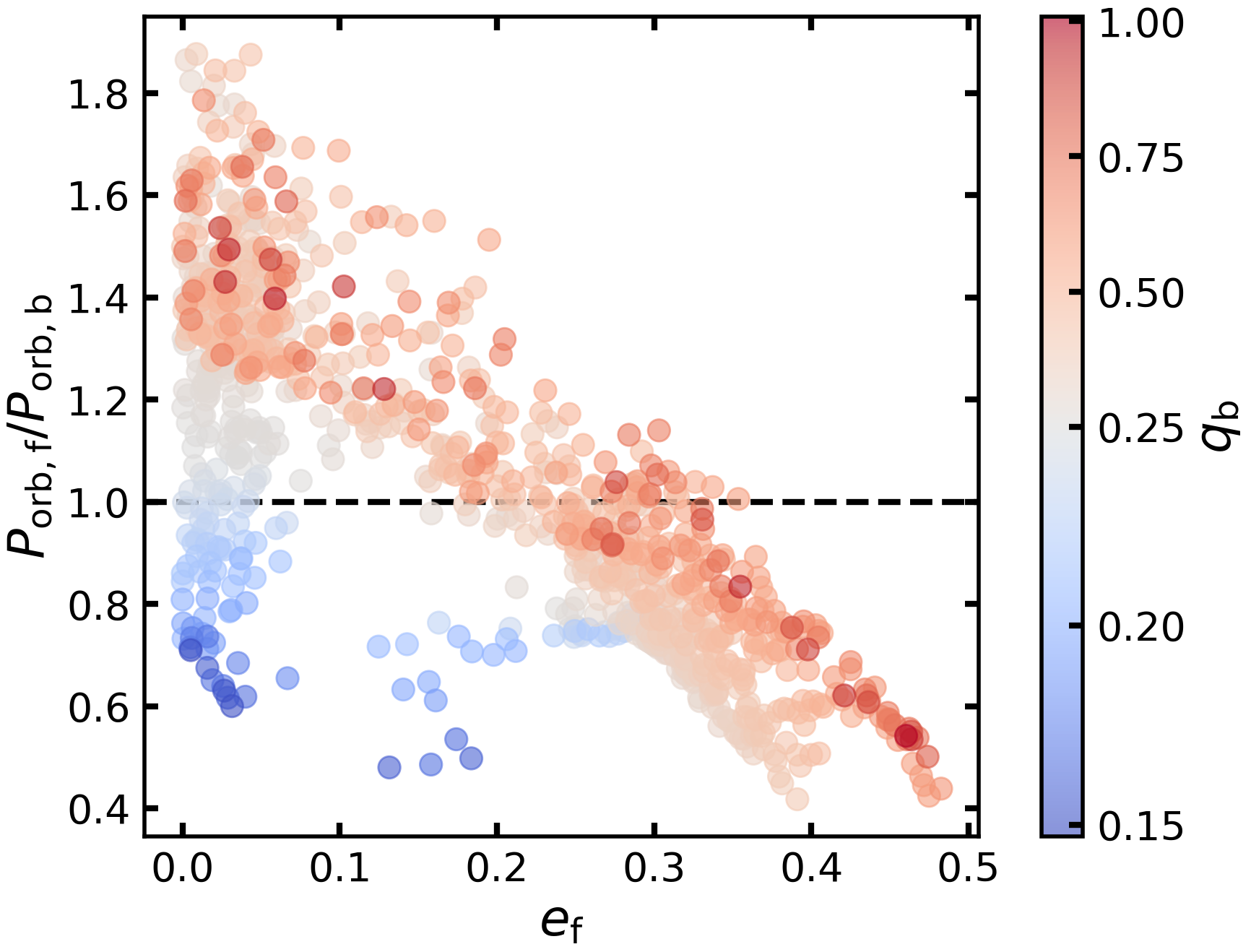}}
    \caption{Relative change in orbital period versus final eccentricity for the model $\#4$ population with the highest $p$-value ($\delta=0.35$ and $e_\mathrm{b,max}=0.07$). The colour scale depicts the mass ratio at the onset of CBD-binary interaction. The black dashed line shows the turning point between orbital widening and shrinkage.}
    \label{figure:model4_Porb_change}
\end{figure}

The relationship between $e_\mathrm{f}$ and the relative change in orbital period, $P_\mathrm{orb,f}/P_\mathrm{orb,b}$, for the model $\#4$ population with the highest $p$-value is shown in Fig.~\ref{figure:model4_Porb_change}, where each system is colour-coded by its value of $q_\mathrm{b}$. The change in orbital period differs greatly depending on whether $q_\mathrm{b}$ is smaller or larger than ${\sim}0.25$. Systems with $q_\mathrm{b}>0.25$ experience orbital widening when $e<0.08$ and subsequent orbital shrinkage when $0.08<e<e_\mathrm{eq}$. Since eccentricity pumping is relatively weak when $e<0.08$, systems with $q_\mathrm{b}>0.25$ significantly increase in $P_\mathrm{orb,f}/P_\mathrm{orb,b}$ until $e_\mathrm{f}\sim0.08$. When $e>0.08$, eccentricity pumping increases in strength and systems with larger $e_\mathrm{f}$ experience an increasingly larger portion of their interaction in this orbital shrinkage regime, which results in the anti-correlation shown in Fig.~\ref{figure:model4_Porb_change}. Systems with $q_\mathrm{b}<0.25$ exhibit orbital shrinkage when $e<e_\mathrm{eq}$. As eccentricity pumping is extremely weak when $e<0.08$, most of these systems reside at $e_\mathrm{f}<0.08$. The few systems with $e_\mathrm{f}>0.08$ do not experience additional orbital shrinkage. 

The general shape shown in Fig.~\ref{figure:model4_Porb_change} remains similar regardless of the values of the free parameters $\delta$ or $\Delta M$ and $e_\mathrm{b}$ or $e_\mathrm{b,max}$ in our models. However, these parameters do influence the range of $P_\mathrm{orb,f}/P_\mathrm{orb,b}$ values of the model systems, which increases with higher $\delta$ or $\Delta M$ values. Higher $e_\mathrm{b}$ or $e_\mathrm{b,max}$ values naturally result in higher $e_\mathrm{f}$ values within the distribution, which retains the same shape.

The change in orbital period depends on the strength of the accretion torques \citep{Siwek2023b}. In the case of SBSG binaries, these torques may be affected by the perceived need for highly inefficient accretion and by the assumptions of the underlying hydrodynamic simulations (see Sect.~\ref{validity of formalism} for further discussion). Furthermore, the implementation of fully inefficient accretion in our model (see Sect.~\ref{disc-binary interaction}) implicitly assumes that the expelled matter carries away specific angular momentum in such a way that the orbital separation and eccentricity evolution via CBD-binary interaction remains unaffected. In Appendix~\ref{angular momentum loss}, we investigate the effect of this assumption and find that it systematically produces shorter orbital period by around $29_{-6}^{+8}\%$ compared to an angular momentum loss prescription in which the matter carries away the specific orbital angular momentum of the binary components via a fast isotropic wind. For these reasons, the results presented above concerning the change in orbital period are relatively uncertain.
 
\subsection{Comparing observed orbital periods to stable mass transfer predictions} \label{stable MT}
In order to compare the eccentricities and orbital periods of the observed systems with those of our model populations, we must assign $a_\mathrm{b}$ values to the modelled systems. Our approach is to assume that they underwent stable mass transfer. This means that the donor star remains approximately equal to its Roche lobe radius throughout mass transfer; in that case, the maximum radius reached by the donor star during mass transfer and the orbital period of the binary at the end of mass transfer are related. As giant stars are expected to exhibit relations between their core mass, radius, and luminosity, SBSG binaries that have undergone stable mass transfer are expected to follow mass-orbital period and mass-luminosity relations \citep[e.g.][]{Webbink1983,Rappaport1995,Lin2011}. 

The orbital periods of observed post-RGB binaries were found by \citet{Moltzer2025} to be broadly consistent with the assumption of stable mass transfer. However, the orbital periods of observed post-AGB binaries are much shorter than expected from stable mass transfer, rendering this assumption inconsistent with the observations \citep[e.g.][]{Nie2012,VanWinckel2025}. Since CBD-binary interaction can shorten the orbital period, we investigated whether stable mass transfer combined with CBD-binary interaction could reproduce the orbital periods of observed post-AGB binaries.

To determine $a_\mathrm{b}$ for each model system, we used the luminosity-orbital period relation for post-RGB binaries derived by \citet{Moltzer2025} from detailed binary stellar evolution models of RGB donors undergoing stable mass transfer. For post-AGB binaries, we estimated the luminosity-orbital period relation using the mass-luminosity relation given in Eq.~\ref{equation:ML} and the mass-orbital period relation from \citet{Rappaport1995}. These relations depend on metallicity, which we assumed to be solar. 

These luminosity-orbital period relations were derived assuming circular orbits during mass transfer, whereas our model systems have non-zero post-mass-transfer eccentricities ($e_\mathrm{b}$). We assumed that mass transfer occurred at periastron of the post-mass-transfer orbit, where $a_\mathrm{per,b}=a_\mathrm{b}(1-e_\mathrm{b})$, as the stars are closest to each other at this point. The orbital period of the system at the onset of CBD-binary interaction is therefore computed using $P_\mathrm{orb,b}=P_\mathrm{orb,circ}/(1-e_\mathrm{b})^{3/2}$, where $P_\mathrm{orb,circ}$ is found using the luminosity-orbital period relation. The orbital period at the end of CBD-binary interaction can then be found using the $P_\mathrm{orb,f}/P_\mathrm{orb,b}$ values computed by our model. 

\begin{figure}
    \resizebox{\hsize}{!}{\includegraphics{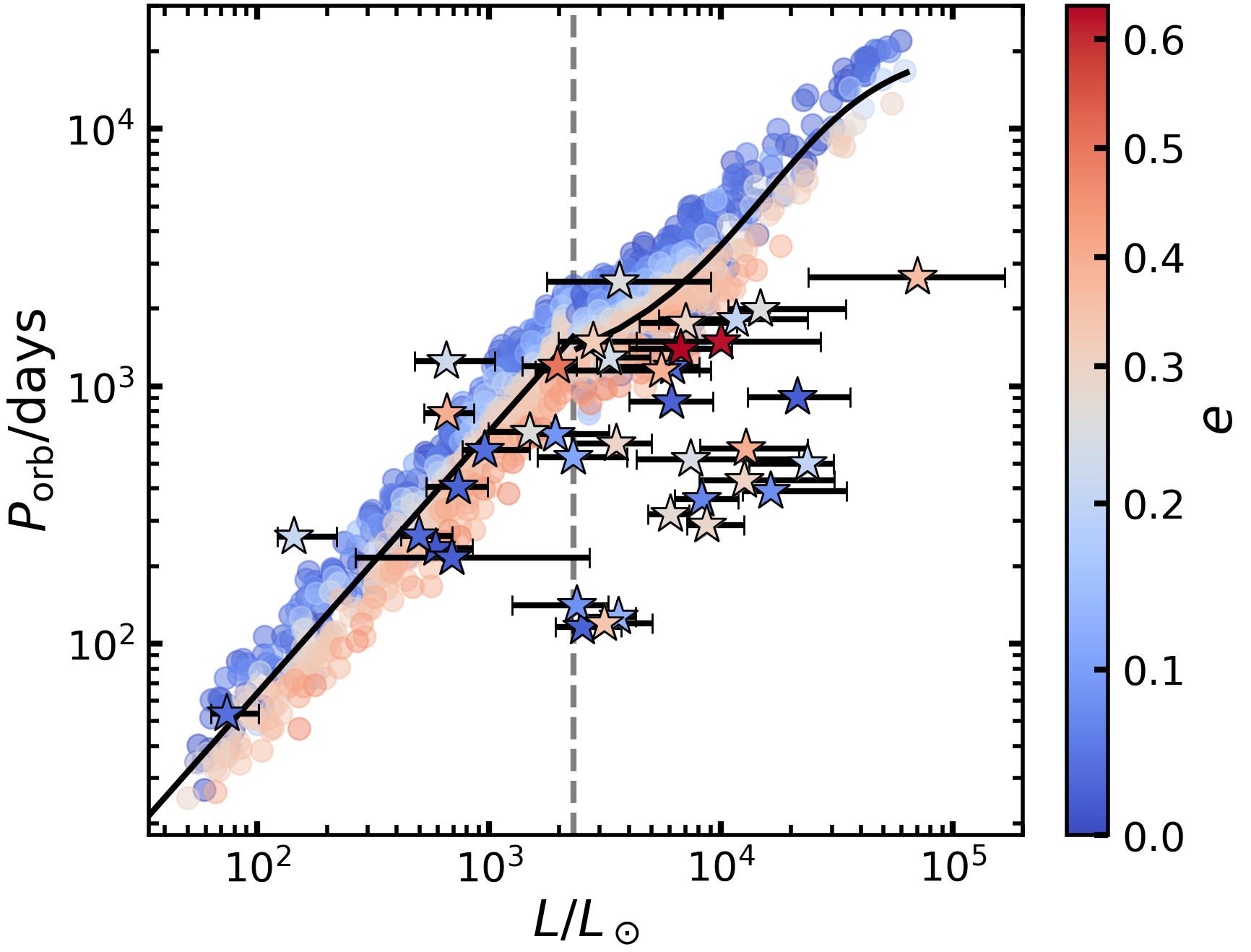}}
    \caption{Luminosity-orbital period diagram of the observed SBSG binaries and the model $\#4$ population with the highest $p$-value ($\delta=0.35$ and $e_\mathrm{b,max}=0.07$), shown by the star- and circle-shaped markers, respectively. The colour scale denotes the eccentricities of the systems. The solid line depicts the luminosity-orbital period relations for post-RGB and post-AGB binaries of solar metallicity. The dashed line corresponds to the RGB-tip luminosity of ${\sim}2300$ $L_\odot$.}
    \label{figure:LPorb}
\end{figure}

The effect of CBD-binary interaction on the luminosity-orbital period relation predicted by stable mass transfer is shown in Fig.~\ref{figure:LPorb}. This process results in systems being spread around the luminosity-orbital period relation. Around $95\%$ of systems above this relation have small eccentricities ($e<0.25$), while around $80\%$ of those below the relation have large eccentricities ($e>0.25$).

The luminosity-orbital period relation for post-RGB binaries presented by \citet{Moltzer2025} is dependent on metallicity, where a lower metallicity results in a shorter orbital period for a given luminosity. The observed post-RGB binaries were found to be broadly consistent with this relation within the expected metallicity range for these systems. However, Fig.~\ref{figure:LPorb} shows that systems undergoing CBD-binary interaction can produce a similar spread in orbital periods for a constant metallicity. In addition, CBD-binary interaction can produce orbital periods that exceed those predicted for the highest expected metallicity (i.e. solar metallicity), which helps reproduce some of the observed orbital periods. 

Of the observed post-AGB binaries in Fig.~\ref{figure:LPorb}, the majority of systems have orbital periods that are up to ten times shorter than predicted by stable mass transfer combined with CBD-binary interaction; only 11 out of 26 systems are marginally consistent with the model population. Using the metal-poor version of the mass-orbital period relation from \citet{Rappaport1995} only shortens the orbital periods by a factor of two. Therefore, we conclude that stable mass transfer followed by CBD-binary interaction cannot explain these shorter orbital periods. This suggests that another process, specifically strong for AGB donors, shortens the orbital period (see Sect.~\ref{discrepancies} for further discussion).

\section{Discussion} \label{discussion}
While our CBD-binary interaction model can reproduce the observed eccentricity distribution of SBSG binaries (see Sect.~\ref{eccentricity}), it has significant limitations, which we discuss in Sect.~\ref{validity of formalism}. Furthermore, several assumptions had to be made in our model that challenge our current understanding of these systems. The most important of these is that the values of the initial parameters $\Delta M$ and $e_\mathrm{b}$ needed to reproduce the observed eccentricity distribution are much higher than those expected from previous studies. In Sects.~\ref{disc masses} and \ref{other_processes}, we discuss the implications of the required $\Delta M$ and $e_\mathrm{b}$ values, respectively. Furthermore, in Sect.~\ref{discrepancies} we speculate on which processes could explain the discrepancies in the eccentricities and orbital periods between our model and the observations.

\subsection{Limitations of CBD-binary interaction model} \label{validity of formalism}
Eccentricity pumping as a result of CBD-binary interaction is governed by the gravitational torques imparted by the CBD onto the binary \citep{Siwek2023b}. Therefore, eccentricity pumping is largely independent of accretion. However, the CBD-binary interaction formalism of \citet{Valli2024} depends on accreted mass rather than time. Consequently, the required $\Delta M$ values serve as a proxy for the necessary timescale for sufficient gravitational torques to be experienced in order to significantly pump the eccentricity. The addition of mass-loss mechanisms (e.g. CBD winds) or other important processes in the context of SBSG binaries \citep[e.g. photoevaporation and macro-structure formation;][]{Oomen2020} could lower the required $\Delta M$ values in our model. This in turn would alleviate the need for highly inefficient accretion (see Sect.~\ref{accretion efficiency}). Nevertheless, even if the required $\Delta M$ values were much smaller, large CBD masses would presumably still be needed to provide the necessary gravitational torque for significant eccentricity pumping.

Although the hydrodynamic simulations of \citet{Siwek2023b} find that the eccentricity is almost always pumped when the initial eccentricity is greater than zero, two similar hydrodynamic simulation studies by \citet{Zrake2021} and \citet{D'Orazio2021} find that systems with initial eccentricities smaller than 0.1 and 0.075, respectively, will experience orbital circularisation instead. As we sample the initial eccentricities of our model systems within the interval $[0,0.1]$, whether eccentricity pumping or orbital circularisation is experienced in this regime will greatly affect our results. If our model employed the findings of \citet{Zrake2021} or \citet{D'Orazio2021}, much larger initial eccentricities would be required to reproduce the observed eccentricities. In contrast to \citet{Siwek2023b}, \citet{Zrake2021} and \citet{D'Orazio2021} only investigated systems with mass ratios equal to unity. Additionally, the equilibrium eccentricity value varies between these CBD-binary interaction studies.

Furthermore, several of the assumptions made in the hydrodynamic simulations by \citet{Siwek2023a,Siwek2023b} need to be assessed in the context of SBSG binaries. Firstly, it is assumed that the radii of the binary components are much smaller than their separation. While this holds true for the main-sequence companion, the SBSG star fills a substantial fraction of its Roche lobe \citep[e.g.][]{Oomen2018}, meaning its radius is larger than that of the sink particle used to model it in the hydrodynamic simulations. In order to mimic a SBSG star in these simulations, a larger sink particle radius is required, which would influence the accretion torques experienced by the binary \citep{Siwek2023b}. This in turn is expected to affect the change in orbital separation due to CBD-binary interaction. However, the eccentricity pumping will remain unaffected as it is governed by gravitational torques from the CBD matter outside of the cavity created by the binary \citep{Siwek2023b}. Therefore, the results in Sect.~\ref{eccentricity} concerning the eccentricity pumping should remain unchanged, while the results in Sect.~\ref{orbital period} regarding the change in orbital period must be viewed with additional uncertainty. 

Secondly, self-gravity of the CBD is ignored in the simulations, since the total mass of the binary is assumed to be much greater than the CBD mass. However, CBDs with masses greater than $1\%$ of the binary mass are expected to be affected by self-gravity \citep{Kratter2016}, which is the case in our modelled systems. Nevertheless, similar hydrodynamic simulations by \citet{Roedig2011} that model self-gravitating CBDs also find that the eccentricity of the binary evolves to an equilibrium value. The equilibrium eccentricity found by \citet{Roedig2011} falls within the interval $[0.6,0.8]$ and is therefore larger than the value of ${\sim}0.5$ found by \citet{Siwek2023b} and compatible with the most eccentric post-AGB binaries observed. 

Thirdly, it is assumed that the CBD is viscously relaxed. This means that the timescale over which the orbital parameters change through CBD-binary interaction has to be longer than the viscous timescale of the inner edge of the CBD, which can be estimated as ${\sim}300$ times the orbital period \citep{Valli2024}. We find that the CBD-binary interaction timescales estimated in Sect.~\ref{accretion efficiency} are always longer than the viscous timescales for post-RGB binaries, but this is not true for the most luminous post-AGB stars (${\gtrsim}1.1\times10^4$ $L_\odot$) when their orbital periods exceed ${\sim}1200$ days. However, only a few systems in the observed post-AGB binary sample fall within this regime.

\subsection{Initial CBD masses} \label{disc masses}
In our CBD-binary interaction model, the amount of accreted mass ($\Delta M$) required to reproduce the observed eccentricity distribution ranges from 0.1 to 1 $M_\odot$. Although, as we discuss in Sect.~\ref{validity of formalism}, the actual accreted masses could be much lower, these $\Delta M$ values still serve as a proxy for the CBD masses required for the systems to experience the necessary gravitational torque to reach high eccentricities. The required CBD masses are much larger than the observed, and hence current, CBD masses of SBSG binaries, which are typically between $10^{-3}-10^{-2}$ $M_\odot$ \citep[e.g.][]{GallardoCava2026}.

However, SBSG binaries already have significantly high eccentricities ($e>0.2$) at the lowest $T_\mathrm{eff}$ values at which they are observed, ranging between $4000-5000$~K. Since most of the evolutionary timescale of SBSG star after mass transfer is spent at even lower $T_\mathrm{eff}$ values than 4000~K \citep[e.g.][]{Moltzer2025}, this suggests that the observed CBDs could be explained as remnants of initially more massive CBDs that pumped the eccentricity of these systems.

We speculate that there is a population of post-mass-transfer progenitor systems with these lower $T_\mathrm{eff}$ values that are actively undergoing CBD-binary interaction. These systems have not yet been found or identified, possibly because they may be obscured by matter expelled during mass transfer before a CBD forms. Furthermore, systems with such low $T_\mathrm{eff}$ values have specifically been excluded from observational searches as they can easily be confused with single RGB and AGB stars \citep[e.g.][]{vanAarle2011,Kamath2014,Kamath2015}.

This proposed population of post-mass-transfer progenitor systems could be related to recently observed samples of post-AGB stars enshrouded by thick dusty envelopes. The first class of these objects, presented by \citet{Khouri2021}, are known as water fountains due to their observed high-velocity water maser emission. These systems are surrounded by slowly expanding material in the shape of a torus, with masses ranging between $0.2-0.8$ $M_\odot$. The water maser emission is believed to result from the interaction between this material and a jet launching perpendicular to the torus. Most of this matter appears to have been ejected within the last 200 years, which \citet{Khouri2021} presume to be due to mass transfer with an unseen binary companion. 

The second class of objects, presented by \citet{Khouri2025}, are called obscured post-AGB stars. This is due to their very red photometric colours, which result from their enshrouding material with masses ranging between $0.1-0.4$ $M_\odot$. Similarly to water fountains, they are observed to exhibit high-velocity emission components and are presumed by \citet{Khouri2025} to have recently undergone a strong phase of mass loss via mass transfer with an unseen binary companion. 

Although water fountains and obscured post-AGB stars have not been confirmed to be in binaries, they resemble the expected post-mass-transfer progenitor systems. While it is unclear what fraction of the recently ejected circumstellar material is in bound orbits around these objects, we speculate that the observed expanding torii could be an intermediate stage in the CBD formation process. If this process occurs within a sufficiently short timescale, it could result in the formation of initially massive CBDs, which are required for our CBD-binary interaction model. If it occurs over a longer timescale, the binary may undergo a similar kind of interaction with the expanding torus. However, such a torus differs greatly from the 2D CBD modelled in our employed CBD-binary interaction formalism, and its outward expansion would affect the required $\Delta M$ values.

\subsection{Additional eccentricity pumping mechanisms} \label{other_processes}
The initial eccentricities required to reproduce the observed eccentricity distribution of SBSG binaries using our CBD-binary interaction model are much larger than the maximum residual eccentricity of $10^{-3}$ expected at the end of a stable mass transfer phase from giant star donors \citep{Phinney1992}. This result is based on the circularisation timescale derived by \citet{Zahn1977} for viscous tidal dissipation in convective envelopes, which predicts that low-mass giant binaries circularise before the onset of RLOF. Although the strength of tidal dissipation is subject to large theoretical uncertainties \citep[e.g.][]{Preece2022,Esseldeurs2024,Dewberry2025}, the circularisation timescale of \citet{Zahn1977} has been extensively tested against observations of pre-RLOF giant binaries \citep[e.g.][]{Verbunt1995,Beck2018,PriceWhelan2018}. We therefore assume that an additional eccentricity pumping mechanism is needed to generate the required initial eccentricities in our model. Four such mechanisms have been described in the literature.

Firstly, the eccentricities of post-mass-transfer binaries have been suggested to result from recoil of the donor star due to asymmetric wind mass loss during its thermally pulsing AGB phase \citep[e.g.][]{Izzard2010,ElBadry2018}. However, this does not account for the required eccentricities of binaries formed from RGB donors, since such stars do not experience comparable episodes of significant wind mass loss. Furthermore, a very large asymmetry is required to achieve a significant eccentricity.

Secondly, the presence of a companion may result in tidally-enhanced wind mass loss, which pumps the eccentricity due to its phase-dependent nature \citep[e.g.][]{BonavicMarinovic2008}. However, as wind mass loss is ineffective for RGB stars, this process cannot account for eccentric post-RGB binaries. The study of \citet{Vos2015} illustrates this, as they found eccentricity pumping via this mechanism to be negligible when forming post-RGB stars with masses close to the helium ignition mass of ${\sim}0.47$~$M_\odot$. 

Thirdly, dynamical interactions with a tertiary companion can pump the eccentricity of a binary system \citep[e.g.][]{Toonen2020}. However, so far no SBSG binary has been observed to be part of a triple system. Furthermore, it is statistically improbable that all eccentric post-mass-transfer binaries are in fact triples \citep{Moe2017}. Nevertheless, the two systems (RU~Cen and V729~Ara) in the observed post-AGB binary sample with eccentricities greater than the equilibrium eccentricity of ${\sim}0.5$ found by \citet{Siwek2023b} could be the result of having an unseen tertiary companion.

Fourthly, phase-dependent RLOF can under some circumstances pump the eccentricity, due to the varying binary separation in an eccentric orbit \citep[e.g.][]{Hamers2019}. A recent study by \citet{Parkosidis2026a} presented a model for phase-dependent RLOF that leads to highly efficient eccentricity pumping in the case of conservative mass transfer. A follow-up study by \citet{Parkosidis2026b} investigating non-conservative mass transfer found that mass loss from the $L_2$ point becomes increasingly likely for higher eccentricities. Furthermore, they found that their model for $L_2$ mass loss results in strong orbital circularisation. Together, these findings suggest that there is a limit to the eccentricity that can be reached by phase-dependent RLOF before $L_2$ mass loss inhibits further eccentricity pumping. 

Of the four eccentricity pumping mechanisms discussed, phase-dependent RLOF appears to be the most promising candidate to supply the required post-mass-transfer eccentricities of at least 0.05 to be able to reproduce the observed eccentricity distribution of SBSG binaries. This is because it is effective for both RGB and AGB donors, and it does not require the statistically improbable presence of numerous tertiary companions. Nevertheless, the other three mechanisms could still play a role for some objects.

\subsection{Explaining the observed eccentricity-orbital period diagram} \label{discrepancies} 
\begin{figure}
    \resizebox{\hsize}{!}{\includegraphics{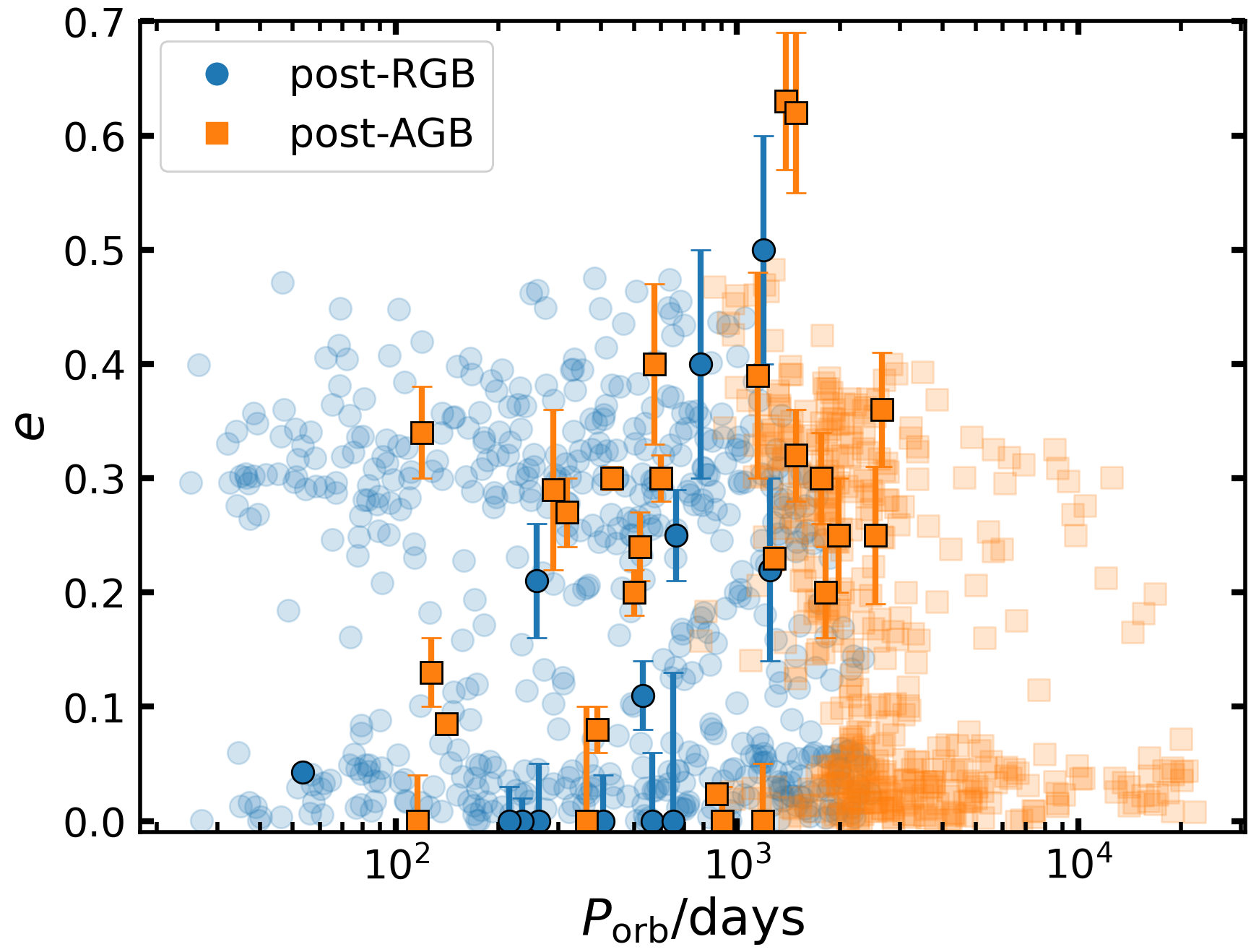}}
    \caption{Eccentricity-orbital period diagram of the observed post-AGB and post-RGB binaries and the model $\#4$ population with the highest $p$-value ($\delta=0.35$ and $e_\mathrm{b,max}=0.07$), shown by the opaque and translucent markers, respectively.}
    \label{figure:e_Porb}
\end{figure}

Fig.~\ref{figure:e_Porb} shows the eccentricities and orbital periods of observed SBSG binaries, compared to the model $\#4$ population with the highest $p$-value for which the orbital periods were computed assuming stable mass transfer (see Sect.~\ref{stable MT}). There are several discrepancies between the model population and the observations, which we speculate can be explained by additional processes missing from our current model.

Evidently, the model post-AGB binaries exhibit the known problem that the orbital periods expected from stable mass transfer are much longer than those of observed post-AGB binaries \citep[e.g.][]{Nie2012,VanWinckel2025}. The process most likely to cause these shorter orbital periods is angular momentum loss via non-conservative mass transfer. In particular, mass loss from the $L_2$ point is an effective mechanism for this \citep[e.g.][]{Soberman1997,Parkosidis2026b}, and it is also the mechanism proposed for forming CBDs \citep[e.g.][]{Frankowski2007,Chen2017}. Therefore, we speculate that the extent of the orbital shrinkage is related to the observed eccentricity, as large amounts of mass lost through $L_2$ would allow for the formation of the initially massive CBDs required for significant eccentricity pumping. We found in Sect.~\ref{pRGB vs pAGB} that our model requires lower initial eccentricities or lower $\Delta M$ values to reproduce the observed eccentricity distribution of post-RGB binaries compared to post-AGB binaries. This could be explained by post-AGB binaries experiencing greater $L_2$ mass loss, resulting in shorter orbital periods and initially more massive CBDs i.e. higher $\Delta M$ values. On the other hand, post-RGB binaries experienced relatively little to no orbital shrinkage compared to stable mass transfer predictions \citep{Moltzer2025}, suggesting that they had initially small CBDs and could therefore not reach high eccentricities. 

The model shown in Fig.~\ref{figure:e_Porb} does not take into account this possible difference between post-RGB and post-AGB binaries. For this model, high-eccentricity post-RGB model systems are predicted at low orbital periods, despite the fact that no such systems have been observed. Another possible explanation for this discrepancy is our assumption that the pre-mass-transfer mass ratio ($q_\mathrm{i}$) must not exceed a fixed critical value ($q_\mathrm{crit}$) in order to ensure stable mass transfer. However, $q_\mathrm{crit}$ actually varies significantly across the viable binary parameter space \citep[e.g.][]{Temmink2023}. In the regime that results in short orbital periods (i.e. smaller donor radii), $q_\mathrm{crit}$ is expected to be much closer to unity; for systems with $q_\mathrm{i}\sim1$, and correspondingly smaller $q_\mathrm{b}$, eccentricity pumping is much weaker. Therefore, implementing an orbital period-dependent $q_\mathrm{crit}$ could enable our model to reproduce the observed trend of increasing eccentricities in post-RGB binaries with longer orbital periods. This trend is observed in many other types of post-interaction low-mass binary system \citep[see e.g.][]{Mathieu2025}, as well as in Gaia astrometric binaries where the companion is probably a white dwarf \citep[e.g.][]{Shahaf2024}; all of these systems also exhibit high eccentricities. These binaries may all have had CBDs shortly after mass transfer, although there is no direct evidence for this. As much more time has elapsed since the mass transfer in these systems than in SBSG binaries, such a CBD would have long since dispersed.

\section{Conclusion} \label{conclusion}
We investigated whether the eccentricity distribution of observed SBSG binaries could be explained by interaction between the binary and the circumbinary disc according to the formalism presented by \citet{Valli2024}. To this end, we generated model populations by sampling the corresponding primary and companion mass distributions derived from observations. The free parameters in this model are the post-mass-transfer eccentricity of each system and the amount of material accreted onto each binary from its CBD. We then statistically compared the eccentricity distributions of the model populations after undergoing CBD-binary interaction with the observed distribution.

We found that CBD-binary interaction can reproduce the observed eccentricity distribution of SBSG binaries, provided that their CBDs had initial masses between 0.1 and 1.0~$M_\odot$ and their initial post-mass-transfer eccentricities can range up to at least 0.05. However, the assumptions required for our model challenge our current understanding of such binaries. 

Firstly, the accretion onto the SBSG star during CBD-binary interaction must be highly inefficient in order to prevent the star from refilling its Roche lobe and initiating another phase of mass transfer. We estimate the maximum accretion efficiency to be between $5\times10^{-5}-0.01$ for post-AGB stars and between $0.007-0.04$ for post-RGB stars. We conclude that the accretion efficiency has a negligible impact on the eccentricity pumping experienced by the model systems. 

Secondly, in order to account for the required amounts of accreted material, the CBDs of SBSG binaries would need to be significantly more massive than current estimates \citep[around $10^{-3}-10^{-2}$~$M_\odot$; e.g.][]{GallardoCava2026}. We speculate that there is a population of post-mass-transfer binaries with lower $T_\mathrm{eff}$ values (below ${\sim}4000$~K) than the observed SBSG binaries, which have CBDs massive enough to facilitate significant eccentricity pumping. We draw parallels between these supposed post-mass-transfer progenitor systems and the water fountains as recently described by \citet{Khouri2021}, as well as obscured post-AGB stars \citep{Khouri2025}. These objects are shrouded in $0.1-0.8$~$M_\odot$ of material recently removed from the observed star, a phenomenon that is tentatively explained by mass transfer via an unseen companion. A systematic search for the proposed post-mass-transfer progenitor systems of SBSG binaries, or establishing the evolutionary link to water fountains and obscured young post-AGB stars, could provide crucial evidence for the validity of our CBD-binary interaction model.

Thirdly, our model requires systems to exhibit much larger eccentricities at the onset of CBD-binary interaction than are theoretically predicted at the end of mass transfer \citep[$\lesssim 10^{-3}$;][]{Phinney1992}. Of the additional eccentricity pumping mechanisms discussed in the literature, phase-dependent RLOF appears to be the most promising candidate for producing substantial post-mass-transfer eccentricities for both post-RGB and post-AGB binaries, although many uncertainties remain. 

Furthermore, we investigated the effect of CBD-binary interaction on the orbital periods of SBSG binaries, assuming they underwent stable mass transfer. We found that, depending on their initial mass ratio and final eccentricity, post-mass-transfer systems undergoing CBD-binary interaction can have their orbits either shrunk or widened. While post-RGB binaries remain consistent with stable mass transfer, as previously found by \citet{Moltzer2025}, the majority of post-AGB binaries continue to have significantly shorter orbital periods than expected. This suggests that these orbits have shrunk due to angular momentum loss through non-conservative mass transfer. One promising candidate for this is mass loss via the $L_2$ point, as this is also a proposed mechanism for forming CBDs \citep[e.g.][]{Frankowski2007,Chen2017}. We speculate that the observed eccentricity is related to the extent of orbital shrinkage, as large amounts of mass lost through $L_2$ would allow for the formation of the initially massive CBDs required for significant eccentricity pumping. In a future study, we will investigate the orbital shrinkage of post-AGB binaries during mass transfer and its effect on their eccentricities. Additionally, since many other types of post-interaction low-mass binary system have similar orbital properties to SBSG binaries \citep[e.g.][]{Shahaf2024,Mathieu2025}, we speculate that they may all have had CBDs with which they interacted shortly after mass transfer.

\section*{Data availability}
Data underlying this article will be shared upon reasonable request to the authors.

\begin{acknowledgements}
The authors thank the anonymous referee for their constructive comments. C.A.S.M. thanks Fiore Stoppa for his advice on using ConTEST. This research made use of NASA’s Astrophysics Data System and the following Python software packages and tools: IPython \citep{Perez2007}, NumPy \citep{Harris2020}, Matplotlib \citep{Hunter2007}, and SciPy \citep{Virtanen2020}.
\end{acknowledgements}

\bibliographystyle{aa}
\bibliography{references}

\begin{appendix}
\section{Post-RGB mass-luminosity relation} \label{pRGB ML}
In order to sample the initial binary properties of our model populations (see Sect.~\ref{population_synthesis}), we determine the mass distribution of SBSG stars in binaries using the luminosities of observed SBSG binaries. Since giant stars exhibit relations between their core mass and luminosity because their dense degenerate cores experience negligible pressure from their extended envelopes \citep[e.g.][]{Refsdal1970}, SBSG stars are expected to exhibit similar relations. As their envelopes have been stripped, the core mass of SBSG stars can be well approximated by their total mass. However, their luminosity will be reduced compared to that of a full-fledged giant star because of their small envelope masses \citep[e.g.][]{Refsdal1970}. Therefore, while the core mass-luminosity relation of giant stars is a useful estimate of the mass-luminosity relation of stripped giant stars, modelling binaries provides a more accurate relation.

For post-AGB stars, we used the single-star models presented by \citet{MillerBertolami2016} to estimate their mass-luminosity relation, since we lack a complete model grid of post-AGB stars formed via stable mass transfer. For post-RGB stars, however, we were able to use the detailed binary evolution models formed via stable mass transfer presented by \citet{Moltzer2025}, covering low- and intermediate-mass donors from the base of the RGB to the first thermal pulse on the AGB. The model grids covered two metallicities ($Z$): approximately solar metallicity ($Z=0.02$) and metal-poor ($Z=0.00142$). As discussed by \citet{Moltzer2025}, post-RGB stars with initial masses of 2.0~$M_{\odot}$ and higher will deviate from the mass-luminosity relation since their cores are still non-degenerate. 

\begin{figure}
    \resizebox{\hsize}{!}{\includegraphics{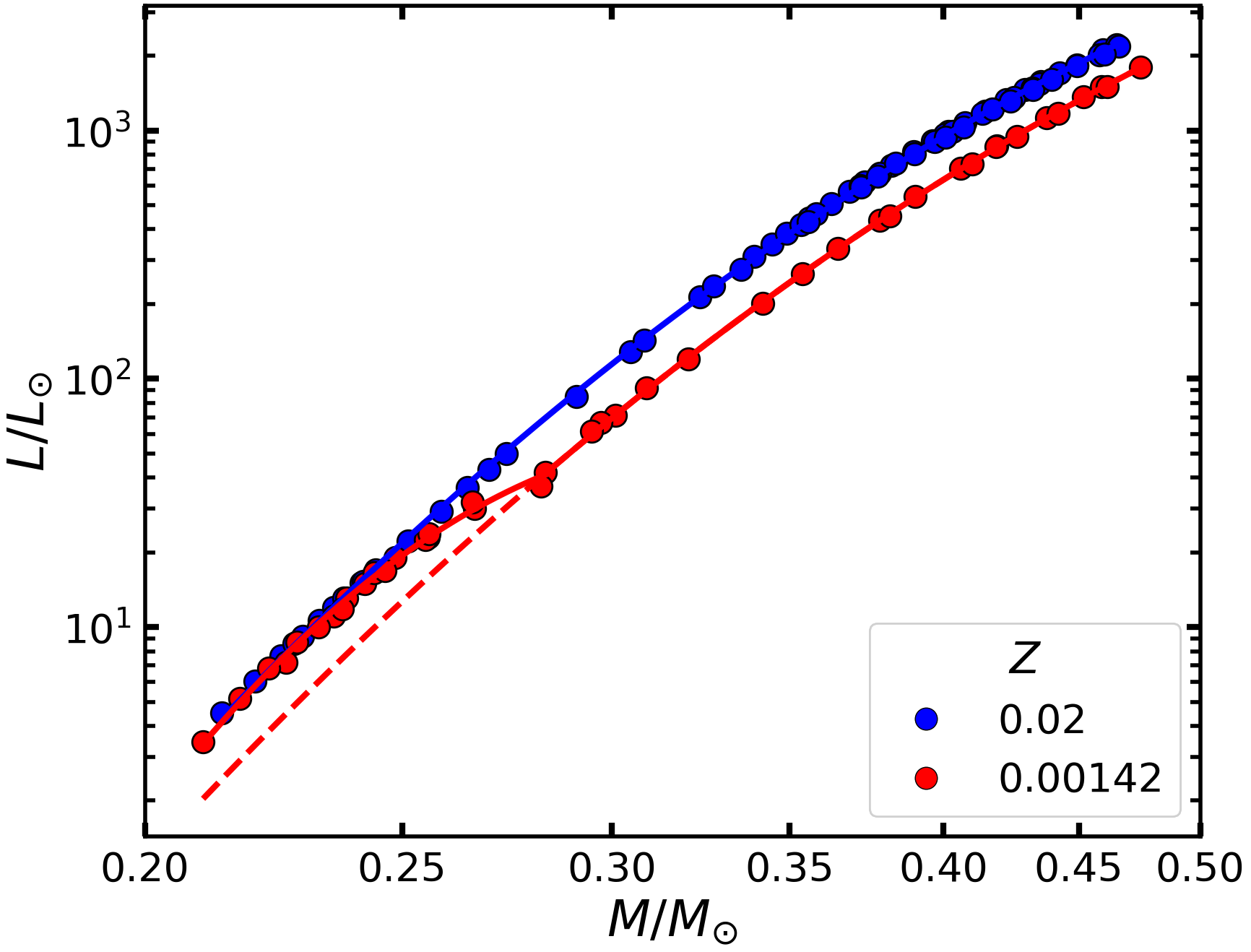}}
    \caption{Luminosity-mass plot of the post-RGB models from \citet{Moltzer2025}, with initial primary masses below $2.0$~$M_{\odot}$ at the end of stable mass transfer, indicated by markers and colour-coded by metallicity. Corresponding fits for the solar and metal-poor grids, given by Eq.~\ref{equation:pRGB ML}, are denoted by solid lines. The dashed line shows the extrapolation of the metal-poor fit for masses above 0.28~$M_\odot$ to lower masses.}
    \label{figure:ML_relations}
\end{figure}

In Fig.~\ref{figure:ML_relations}, the masses and luminosities of the model post-RGB stars with initial masses less than 2.0~$M_{\odot}$ are shown at the end of mass transfer, which was defined as the mass transfer rate dropping below $10^{-12}$~$M_\odot/\mathrm{year}$. Unlike the post-AGB star models from \citet{MillerBertolami2016}, the luminosities of the post-RGB star models exhibit a clear metallicity dependence. Since the luminosity produced by hydrogen-shell burning depends strongly on the mean molecular weight of the burning shell \citep[e.g.][]{Kippenhahn1981}, a higher metallicity results in a higher luminosity for a given mass. For simplicity, we used the solar metallicity models to compute the masses of observed post-RGB stars from their luminosities in Sect.~\ref{population_synthesis}.

We found that the mass-luminosity relations of the post-RGB models shown in Fig.~\ref{figure:ML_relations} can be well fitted by
\begin{equation}
    \log(L/L_\odot) = b_0 + b_1 \log(M/M_{\odot}) + b_2 \log^2(M/M_{\odot}),
    \label{equation:pRGB ML}
\end{equation}
which has a similar form to the core mass-luminosity relation for RGB stars by \citet{Webbink1983}. The coefficients in Eq.~\ref{equation:pRGB ML} for the solar metallicity models were found to be 
\begin{equation*}
Z=0.02:
\begin{cases}
    b_0 = 4.08\pm0.03, \\
    b_1 = -0.7\pm0.1, \\
    b_2 = -8.7\pm0.1, \\
\end{cases}
0.21<M/M_{\odot}<0.47. \\
\end{equation*}
The metal-poor models had to be fitted as a piecewise relation, with the coefficients in Eq.~\ref{equation:pRGB ML} being
\begin{equation*}
Z=0.00142:
\begin{cases}
    b_0 = -5.5\pm1.2, \\
    b_1 = -30\pm4, \\
    b_2 = -31\pm3, \\
\end{cases}
0.21<M/M_{\odot}<0.28, \\
\end{equation*}
\begin{equation*}
Z=0.00142:
\begin{cases}
    b_0 = 4.2\pm0.1, \\
    b_1 = 0.2\pm0.5, \\
    b_2 = -8.1\pm0.5, \\
\end{cases}
0.28<M/M_{\odot}<0.47. \\
\end{equation*}
The change in the mass-luminosity relation at 0.28~$M_\odot$ of these metal-poor models is related to the H-discontinuity resulting from the first dredge-up. This occurs when the convective envelope mixes H-burning ashes from deep inside the star during the RGB phase, leaving behind a discontinuity in the hydrogen abundance profile at its deepest extent. When the H-burning shell reaches this discontinuity, its luminosity decreases due to the lower mean molecular weight resulting from the higher hydrogen abundance \citep[e.g.][]{Christensen-Dalsgaard2015}. Since more metal-rich stars have deeper convective envelopes \citep[e.g.][]{Angelou2015}, their burning shells encounter the H-discontinuity at a lower core mass. This explains why all solar metallicity post-RGB models experienced the H-discontinuity, but the metal-poor models below 0.28~$M_\odot$ did not, as the layer containing the H-discontinuity is stripped during mass transfer before the burning shell can reach it.

\section{Comparing eccentricity distributions using \texttt{contest$\_$dens}} \label{ConTEST details}
In order to compare the eccentricity distribution of our model with that observed for SBSG binaries, we employed the statistical consistency test \texttt{contest$\_$dens}\footnote{The corresponding Python function is publicly available at \url{https://github.com/FiorenSt/ConTEST}.} presented by \citet{Stoppa2023}. This method tests whether the model hypothesised for being responsible for the observed eccentricity distribution (i.e. the null hypothesis) can be rejected or not. Firstly, by approximating the probability density functions of the observed and model distributions ($f$ and $g$, respectively) using kernel density estimation, the distance between $f$ and $g$ is computed (i.e. the test statistic). Secondly, by simulating $K$ probability density functions ($f_k$) from $g$ via $n$ samplings, $K$ distances between $f_k$ and $g$ are computed. Thirdly, a $p$-value is computed by determining the proportion of the $K$ distances that are larger than the test statistic. Lastly, the $p$-value is compared with a significance level $\alpha$, and model populations with $p$-values greater than $\alpha=0.05$ will be labelled as statistically significant (i.e. the null hypothesis is not rejected). We use $K=1000$ and $n=1000$ as is done by \citet{Stoppa2023}.

To ensure that the probability density functions have the required domain for $e$ of $[0,1]$, we transformed the data and model eccentricities using the logit function, defined as $\mathrm{logit}(x) = \ln \left( \frac{x}{1-x} \right)$. However, the domain of this function is $\{0,1\}$, and some of the observed SBSG binaries have an eccentricity of 0. To circumvent this issue, we assumed that the eccentricities of these systems are each normally distributed with a mean of 0 and a standard deviation equal to the observed standard error, truncated between 0 and 1. We then compute the corresponding non-zero median values (see Appendix~\ref{truncated median}).

Although the utilised consistency test \texttt{contest$\_$dens} can compare multiple parameters simultaneously, we chose to limit the comparison to eccentricities only, rather than utilising both eccentricities and luminosities. This is because the uncertainties on the luminosities of SBSG binaries are significantly larger than those on the eccentricities, and this test does not take this into account.

\section{Medians of truncated normal distributions} \label{truncated median}
Consider a normal distribution of a variable $x$ with a mean value $\mu$ and a standard deviation $\sigma$, bounded by the interval $a \leq x \leq b$. The cumulative distribution function $F(x)$ of this truncated normal distribution is defined as
\begin{equation}
   F(x) = \frac{\Phi\left(\frac{x-\mu}{\sigma}\right) - \Phi\left(\frac{a-\mu}{\sigma}\right)}{\Phi\left(\frac{b-\mu}{\sigma}\right) - \Phi\left(\frac{a-\mu}{\sigma}\right)}.
\end{equation}
Here, $\Phi$ corresponds to the cumulative distribution function of the standard normal distribution given by
\begin{equation}
   \Phi(x) = \frac{1}{2} \left( 1+ \mathrm{erf} \left( \frac{x}{\sqrt{2}} \right) \right),
\end{equation}
where $\mathrm{erf}$ is the error function. As the median $\tilde{x}$ is defined as $F(\tilde{x}) = 1/2$, $\tilde{x}$ for this truncated normal distribution can be expressed as
\begin{equation}
   \tilde{x} = \mu + \sigma \Phi^{-1}\left( \frac{\Phi\left(\frac{a-\mu}{\sigma}\right) + \Phi\left(\frac{b-\mu}{\sigma}\right)}{2}\right).
    \label{equation:median_trunc}
\end{equation}

The eccentricities ($e$) of SBSG binaries are observationally measured ($\mu_e$) with a given uncertainty ($\sigma_e$). As the eccentricity of an orbit is, by definition, bounded between 0 and 1, the uncertainty interval of an observed eccentricity can exceed these bounds. Assuming that the eccentricity of each SBSG binary is normally distributed, where $\mu_e$ is the mean value and $\sigma_e$ is the standard deviation, we use $\mu=\mu_e$, $\sigma=\sigma_e$, $a=0$, and $b=1$ in Eq.~\ref{equation:median_trunc} to compute the median value $\tilde{e}$ of this truncated normal distribution. We chose to use the median instead of the mean as the most likely value of $e$ because the mean is sensitive to outliers for a skewed distribution such as this. For the majority of observed systems, $\tilde{e}$ is equal to $\mu_e$; only the ten binaries with $\mu_e=0$ have distributions that are truncated, resulting in non-zero $\tilde{e}$ values.

\section{Comparing mass function distributions} \label{mass function}
In Sect.~\ref{model3_results}, we constrained the pre-mass-transfer mass ratio ($q_\mathrm{i}$) of each system to $1 \leq q_\mathrm{i} \leq 2$ in our model $\#3$ to ensure that these systems were likely formed via stable mass transfer. As a result, the distribution of the companion masses ($M_2$) of the model populations differs from the companion mass distribution of \citet{Oomen2018}, from which these $M_2$ values were originally sampled (see Fig.~\ref{figure:model3_Mcomp_comparison}). We performed a statistical test to determine whether the $M_2$ distribution of the model $\#3$ populations is distinguishable from the observed distribution of \citet{Oomen2018}.

The companion mass distribution of \citet{Oomen2018} was constructed to reproduce the mass functions of 33 SBSG binaries (see their Table 2). The mass function $f(m)$ is an observable of a spectroscopic binary system which is related to the masses of its components 
\begin{equation}
   f(m) = \frac{M_2^3}{(M_1+ M_2)^2}\sin^3(i),
    \label{equation:mass_function}
\end{equation}
where $i$ is the inclination angle, i.e. the angle between the orbital axis of the binary system and the line of sight. Therefore, we compared the observed mass function distribution to that of the model population. We computed the mass functions of the model population using Eq.~\ref{equation:mass_function}, uniformly sampling $\cos(i)$ to account for the random orientation of the orbits. Similarly to \citet{Oomen2018}, we imposed a maximum value of $75^{\circ}$ for $i$ as we do not expect to observe SBSG binaries edge-on due to their CBDs. 

\begin{figure}[h!]
    \resizebox{\hsize}{!}{\includegraphics{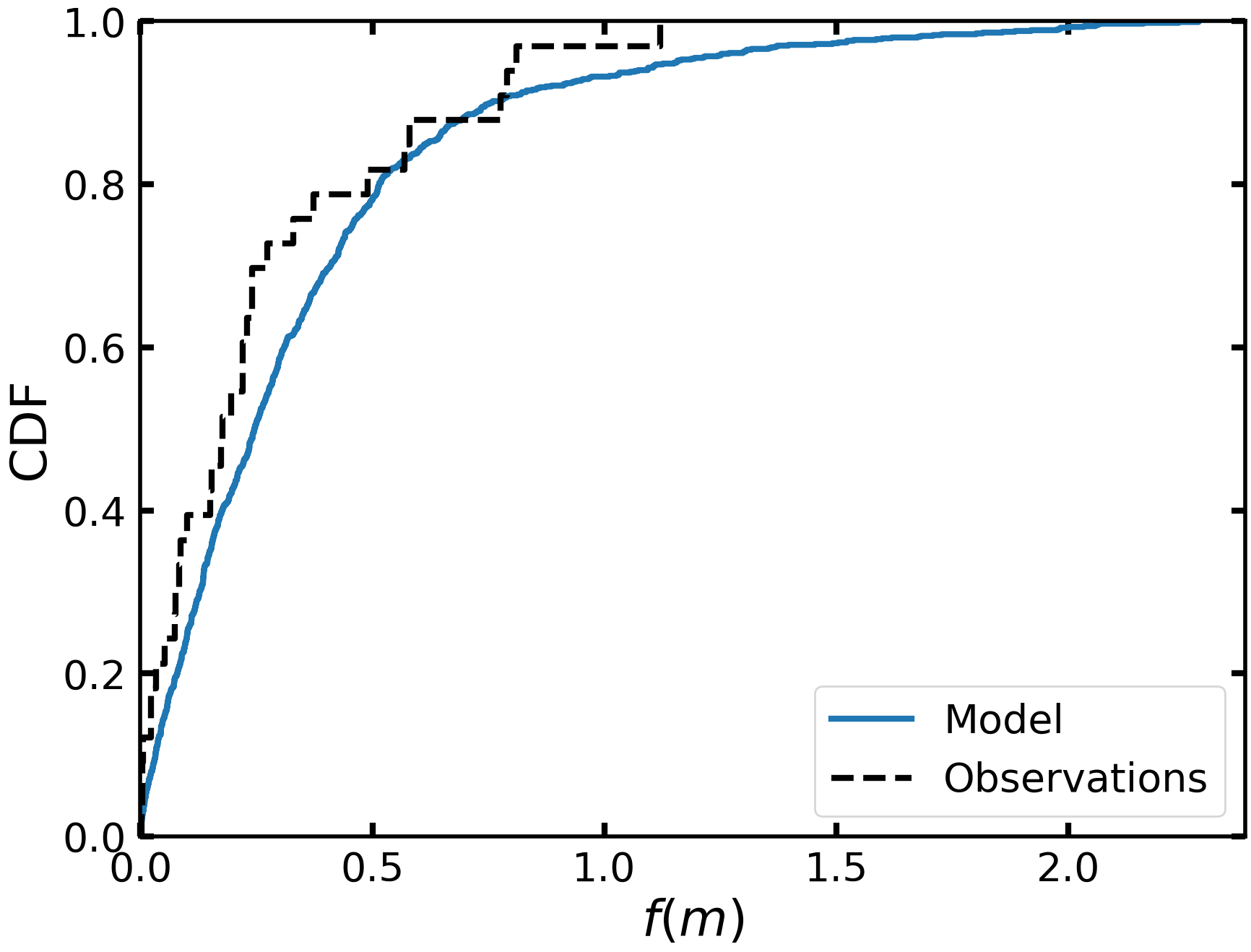}}
    \caption{Cumulative mass function distributions of the observed systems used by \citet{Oomen2018} to construct the $M_2$ distribution, and of a model population with the imposed constraint of $1 \leq q_\mathrm{i} \leq 2$ sampled from this $M_2$ distribution of \citet{Oomen2018}. The mass functions for the model population were computed using random uniform sampling over the cosine of the inclination angle, in order to account for the random orientation of the orbits.}
    \label{figure:model3_fm_comparison}
\end{figure}

The resulting mass function distribution of the model population is compared to the observed distribution in Fig.~\ref{figure:model3_fm_comparison}. Using the statistical consistency test \texttt{contest$\_$dens} (see Appendix~\ref{ConTEST details} for further details), we found a statistically significant $p$-value of ${\sim}0.4$. This indicates that, despite the differences in the $M_2$ distributions, the model distribution cannot be ruled out as being equivalent to the observed one of \citet{Oomen2018}. This is because a significant change in the $M_2$ distribution only has a small effect on the mass function distribution.

\section{Effect of accretion efficiency on the eccentricity distribution} \label{effect on orbital parameters}
In our CBD-binary interaction model, we assumed that the accretion onto the stars themselves is fully inefficient, since only a small amount of matter is needed for the SBSG stars to regain their envelopes and to start filling their Roche lobes again (see Sect.~\ref{accretion efficiency}). As the eccentricity pumping is primarily caused by gravitational torques \citep{Siwek2023b}, which therefore remain unaffected by changes in accretion, we expect this assumption to have a negligible effect on our model eccentricity distributions. In order to verify this, we compare models with fully inefficient and fully efficient accretion, in which $q$ evolves according to the findings of \citet{Siwek2023b}.

In the case of fully efficient accretion, the $M_1$ and $M_2$ distributions will shift towards higher values as $q$ evolves towards unity. As $M_1$ is smaller than $M_2$ in almost all of our model systems, the accretion affects the $M_1$ distribution the most. Although $\Delta M$ is not equally split between the two binary components, the difference in the amount of accreted mass between the two components is never greater than $6\%$. Nevertheless, this substantially shifts the $M_1$ distribution towards larger masses.

\begin{figure}[h!]
    \resizebox{\hsize}{!}{\includegraphics{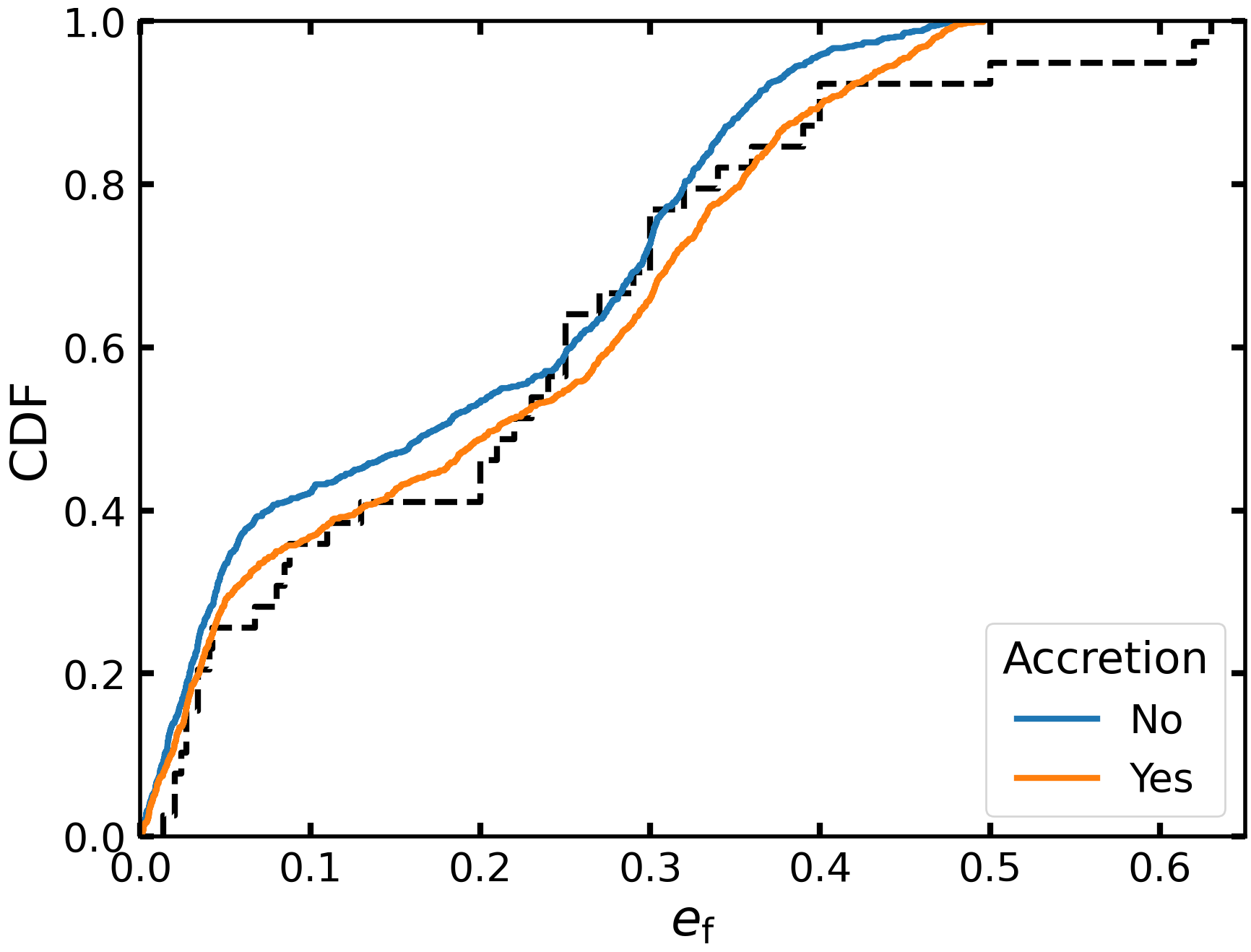}}
    \caption{Cumulative eccentricity distribution of the model $\#4$ population with the highest $p$-value ($\delta=0.35$ and $e_\mathrm{b,max}=0.07$), both with and without accretion onto the stars themselves. The dashed black line denotes the observed eccentricity distribution of SBSG binaries, where the eccentricities are defined as the median values of their corresponding truncated normal distributions (see Appendix~\ref{truncated median}).}
    \label{figure:model4_accretion}
\end{figure}

We compare the eccentricity distribution of the model $\#4$ population with the highest $p$-value ($\delta=0.35$ and $e_\mathrm{b,max}=0.07$) with that of the same model population for which accretion is fully efficient, as shown in Fig.~\ref{figure:model4_accretion}. There are two noticeable differences between these eccentricity distributions: fully efficient accretion results in fewer systems with $e_\mathrm{f}<0.08$ and more systems with $e_\mathrm{f}>0.4$. These differences arise directly from the lack of $q$ evolution in the fully inefficient scenario. Eccentricity pumping is substantially stronger for systems with $q>0.4$ when $e<0.08$, and for systems with $q>0.8$ when $e>0.08$. This results in fewer low-eccentricity systems and more high-eccentricity systems, respectively, when $q$ evolution allows systems to reach these stronger eccentricity pumping regimes during CBD-binary interaction. Nevertheless, the impact of the accretion efficiency on the final eccentricity distribution is small and the fully efficient scenario still yields a highly significant $p$-value when compared to the observations. We conclude that the observed eccentricity distribution does not allow for the accretion efficiency to be constrained.

\section{Angular momentum loss prescription} \label{angular momentum loss}
As described in Sect.~\ref{disc-binary interaction}, our model assumes that accretion from the circumbinary disc onto the binary is fully inefficient. We implement this by disregarding the mass ratio evolution presented by \citet{Siwek2023a}. While this approach was chosen for simplicity, it does implicitly assume that the expelled matter carries away specific angular momentum such that the orbital separation and eccentricity evolution remains unaffected. Here we investigate the effects of our assumed angular momentum loss prescription on the results in Sect.~\ref{orbital period}, by comparing it with a standard prescription from the literature.

The orbital angular momentum of a binary system can be expressed as
\begin{equation}
   J_\mathrm{b} = M_\mathrm{p} M_\mathrm{s} \left(\frac{Ga(1-e^2)}{M_\mathrm{b}}\right)^{1/2} = \frac{q}{(1+q)^2} \left[ GM_\mathrm{b}^3a(1-e^2) \right]^{1/2},
   \label{equation:J}
\end{equation}
where $M_\mathrm{b}=M_\mathrm{p}+M_\mathrm{s}$ and $q = M_\mathrm{s}/M_\mathrm{p} \leqslant 1$. By differentiating this expression, the change in orbital angular momentum of the binary can be expressed as
\begin{equation}
   \frac{dJ_\mathrm{b}}{J_\mathrm{b}} = \frac{3}{2}\frac{dM_\mathrm{b}}{M_\mathrm{b}}  + \frac{1-q}{1+q}\frac{dq}{q} + \frac{1}{2}\frac{da}{a} - \frac{e\,de}{1-e^2}.
   \label{equation:dJ}
\end{equation}
In the presence of mass loss from the binary, e.g. due to inefficient accretion, the angular momentum taken away per unit mass lost can be defined as
\begin{equation}
   j_\mathrm{loss} = \frac{dJ}{dM} \equiv \gamma \frac{J_\mathrm{b}}{M_\mathrm{b}},
   \label{equation:jloss}
\end{equation}
where $\gamma$ is a convenient quantity that parameterises $j_\mathrm{loss}$ in terms of the specific angular momentum of the binary.

In the scenario of fully efficient accretion (i.e. when $\epsilon=1$), as modelled in the simulations of \citet{Siwek2023a,Siwek2023b}, the changes in $e$, $a$ and $q$ are given by Eqs.~\ref{equation:da}-\ref{equation:dq}, and the binary mass increases by the amount of mass accreted from the CBD, so that $dM_\mathrm{b} = dM$. Eq.~\ref{equation:dJ} then gives a change in orbital angular momentum equal to
\begin{equation}
   \left( \frac{dJ_\mathrm{b}}{J_\mathrm{b}} \right)_{\epsilon=1} = \left[ \frac{3}{2} + \frac{1-q}{1+q}\frac{f_q(e,q)}{q} + \frac{1}{2}f_a(e,q) - \frac{e}{1-e^2}f_e(e,q) \right] \frac{dM}{M_\mathrm{b}}.
   \label{equation:dJdM}
\end{equation}
In the fully inefficient accretion scenario (i.e. when $\epsilon=0$), the binary mass and mass ratio do not evolve, so that $dM_\mathrm{b}=0$ and $dq=0$ in Eq.~\ref{equation:dJ}. However, the $da$ and $de$ terms can still be written in terms of $dM$ using Eqs.~\ref{equation:da} and \ref{equation:de}, respectively, since here $dM$ represents the mass flowing from the CBD towards the binary and subsequently being ejected. Therefore, Eq.~\ref{equation:dJ} for our model with $\epsilon=0$ can be expressed as
\begin{equation}
   \left( \frac{dJ_\mathrm{b}}{J_\mathrm{b}} \right)_{\epsilon=0} = \left[ \frac{1}{2}f_a(e,q) - \frac{e}{1-e^2}f_e(e,q) \right] \frac{dM}{M_\mathrm{b}}.
\end{equation}
The difference between the cases with $\epsilon=1$ and $\epsilon=0$ represents the angular momentum taken away by the mass $dM$ in our model with fully inefficient accretion. Using Eq.~\ref{equation:jloss}, this yields a specific angular momentum loss parameter equal to
\begin{equation}
   \gamma = \frac{3}{2} + \frac{1-q}{1+q}\frac{f_q(e,q)}{q}.
   \label{equation:gamma}
\end{equation}

We can compare this to theoretical angular momentum loss prescriptions. Several standard angular momentum loss modes have been described in the literature \citep[e.g.][]{Soberman1997}. In the context of our model, the two most applicable modes are those in which the mass accreted by the primary and secondary is lost via a fast isotropic wind that carries away the specific orbital angular momentum of the binary components in their relative orbit around the centre of mass. For these modes, $\gamma$ is equal to $q$ and $1/q$ for the wind from the primary and secondary, respectively \citep[for the derivations, see e.g.][]{Parkosidis2026b}. We assume that the fraction of mass lost from each binary component equals the fraction of mass it accretes in the model of \citet{Valli2024}. These fractions can be defined as
\begin{equation}
   dM_\mathrm{p} = f_\mathrm{p}dM, 
   \quad\mathrm{and}\quad
   dM_\mathrm{s} = f_\mathrm{s}dM,
   \label{equation:fdM}
\end{equation}
where $f_\mathrm{p}+f_\mathrm{s}=1$. Using Eq.~\ref{equation:dq}, these fractions can be written in terms of $f_q(e,q)$ as
\begin{equation}
   f_\mathrm{p} = \frac{1}{1+q} - \frac{f_q(e,q)}{(1+q)^2},
   \quad\mathrm{and}\quad
   f_\mathrm{s} = \frac{q}{1+q} + \frac{f_q(e,q)}{(1+q)^2}.
   \label{equation:f}
\end{equation}
Therefore, under the assumption of isotropic mass loss from each binary component, we expect the specific angular momentum loss parameter to be given by
\begin{equation}
   \gamma^{\prime} = q f_\mathrm{p} + \frac{1}{q} f_\mathrm{s} = 1 + \frac{1-q}{1+q}\frac{f_q(e,q)}{q}.
\end{equation}

\begin{figure}
    \resizebox{\hsize}{!}{\includegraphics{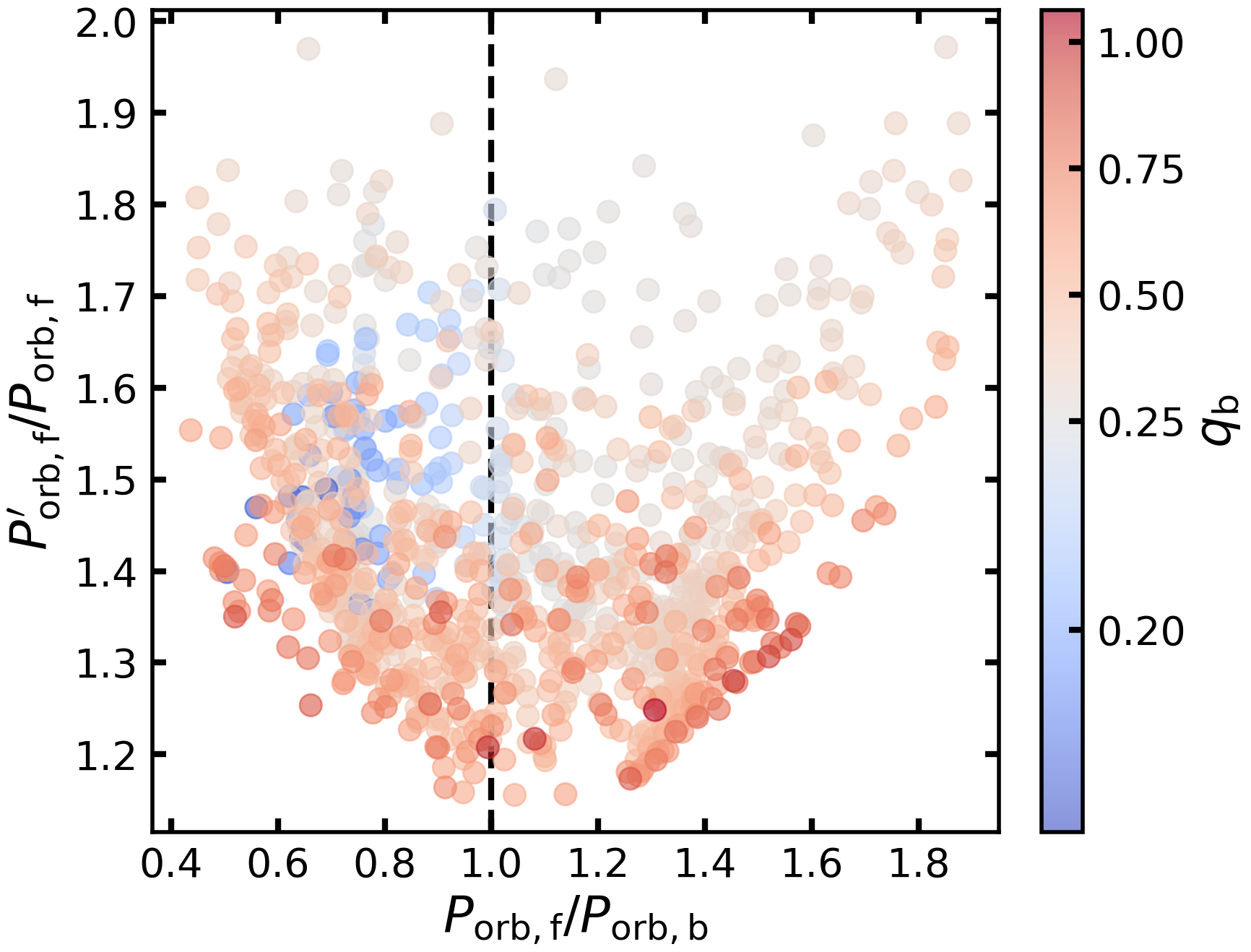}}
    \caption{Relative difference in the final orbital period between the isotropic mass loss case and our assumed angular momentum loss prescription versus relative change in orbital period for the model $\#4$ population with the highest $p$-value ($\delta=0.35$ and $e_\mathrm{b,max}=0.07$). The colour scale depicts the mass ratio at the onset of CBD-binary interaction. The black dashed line shows the turning point between orbital widening and shrinkage.}
    \label{figure:model4_Porb_change_comparison}
\end{figure}

We find that the angular momentum loss expected under isotropic mass loss differs from the angular momentum loss assumed in our model by a constant value of $\gamma^\prime - \gamma = -1/2$, which means that angular momentum loss is always greater in our model. We demonstrate the effect of this difference on the orbital periods predicted by our model. Using Eq.~\ref{equation:jloss} and integrating over $dM$, the difference in final orbital angular momentum between the two cases after a mass $\Delta M$ has been accreted from the CBD onto the binary amounts to
\begin{equation}
\ln J^\prime_\mathrm{b,f} - \ln J_\mathrm{b,f} = \int_{\Delta M} -(\gamma^\prime - \gamma) \frac{dM}{M_\mathrm{b}} = \frac{1}{2} \frac{\Delta M}{M_\mathrm{b}},
\end{equation}
since $M_\mathrm{b}$ is constant in our fully inefficient accretion model. We assume that the difference in angular momentum loss affects only the orbital period of the binary and not its eccentricity, since eccentricity pumping is governed by gravitational torques rather than by those resulting from accretion \citep{Siwek2023b}. Since $J_\mathrm{b} \propto a^{1/2} \propto P_\mathrm{orb}^{1/3}$, the relative difference in the final orbital period between the isotropic mass loss case and our assumed angular momentum loss prescription can be expressed as
\begin{equation}
   \frac{P_\mathrm{orb,f}^{\prime}}{P_\mathrm{orb,f}} = \exp \left( \frac{3}{2} \frac{\Delta M}{M_\mathrm{b}} \right).
\end{equation}
We show $P_\mathrm{orb,f}^{\prime}/P_\mathrm{orb,f}$ in Fig.~\ref{figure:model4_Porb_change_comparison} for the model $\#4$ population with the highest $p$-value ($\delta=0.35$ and $e_\mathrm{b,max}=0.07$). The isotropic mass loss prescription would result in an increase in the orbital period ranging between $16\%$ and $97\%$, with a median increase of $41\%$ and a spread around this value of $-12\%$ and $19\%$ for the 16th and 84th percentiles, respectively.

We conclude that, although the non-standard angular momentum loss prescription assumed in our model predicts systematically shorter orbital periods than a prescription based on isotropic mass loss, it does not change the overall conclusion we draw in Sect.~\ref{stable MT}: the orbital periods of observed post-RGB binaries can be explained by stable mass transfer, whereas the orbital periods of most observed post-AGB binaries are much shorter than predicted.

\end{appendix}

\end{document}